%% file: effRDf-paper.tex
\documentclass[twocolumn,aps,prd,floatfix,preprintnumbers,a4paper,nofootinbib,superscriptaddress]{revtex4-1}
\usepackage{graphicx}				
\usepackage{placeins}
\usepackage{amssymb}
\usepackage{amsmath}
\usepackage[dvipsnames]{xcolor}
\usepackage{verbatim}
\usepackage{rotating}
\usepackage{comment}
\usepackage[colorlinks]{hyperref}

\begin{document}

\input{macro}

\newcommand{\Zurich}{Physik-Institut, Universit\"at Z\"urich, Winterthurerstrasse 190, 8057 Z\"urich, Switzerland}
\newcommand{\MIT}{MIT-Kavli Institute for Astrophysics and Space Research and LIGO Laboratory, 77 Massachusetts Avenue, 37-664H, Cambridge, MA 02139, USA}
\newcommand{\Amsterdam}{Institute for High-Energy Physics, University of Amsterdam, Science Park 904, 1098 XH Amsterdam, The Netherlands}
\newcommand{\Cardiff}{School of Physics and Astronomy, Cardiff University, Cardiff, CF24 3AA, United Kingdom}
\newcommand{\KCL}{King's  College  London,  Strand,  London  WC2R  2LS,  United Kingdom}

\title{Ringdown frequencies in black holes formed from precessing black-hole binaries}

\author{Eleanor Hamilton} \affiliation{\Zurich}
\author{Lionel London} \affiliation{\KCL} \affiliation{\Amsterdam} \affiliation{\MIT} 
\author{Mark Hannam} \affiliation{\Cardiff}

\begin{abstract}
We present a simple formula for the effective ringdown frequencies of the gravitational-wave signal of a precessing black-hole binary in the 
co-precessing frame. 
This formula requires only knowledge of the quasi-normal mode frequencies of the system and the value of the precession angle $\beta$ during ringdown.
Such a formula will be useful in modelling precessing systems.
We also provide a comprehensive description of the oscillations in the ringdown frequency in an inertial frame where the spin of the final black hole is in the $z$-direction. 
These oscillations arise due to the superposition of the prograde and retrograde frequencies.
Our understanding of these oscillations can be used to extract the ratio of the amplitudes of the prograde and retrograde frequencies from numerical data.
Alternatively, knowledge of this ratio of the amplitudes can be used to produce a simple model of the time domain oscillations in the ringdown frequency.
\end{abstract}

\maketitle

\section{Introduction}

To date the LIGO~\cite{LIGOScientific:2014pky} and Virgo~\cite{VIRGO:2014yos} detectors, as part of the LIGO-Virgo-Kagra collaboration, have detected 90 \gw{} signals; 83 of which are confidently considered to come from \bbh{} mergers~\cite{LIGOScientific:2018mvr,LIGOScientific:2020ibl,LIGOScientific:2021djp}.
These have included evidence of stellar mass \bh{s}, intermediate mass \bh{s} and those with masses in the Pair-instability Mass Gap~\cite{LIGOScientific:2020iuh,LIGOScientific:2020ufj}.
Binaries with mass ratios up to $\sim10$~\cite{LIGOScientific:2020zkf} and those with evidence of spinning black 
holes~\cite{LIGOScientific:2016sjg,LIGOScientific:2020stg,LIGOScientific:2020ibl,LIGOScientific:2021djp}
have also been observed.
Analysis of these signals requires accurate models of \gw{} signals from \cbc{s} for use in accurate parameter estimation with these detections. A detailed, comprehensive understanding of the nature of the compact objects comprising binaries emitting \gw{s} enables us to make confident astrophysical inferences as to the population of such objects in the Universe~\cite{LIGOScientific:2021psn}.
Highly accurate models can also be used in precision tests of \gr{}~\cite{LIGOScientific:2016lio, LIGOScientific:2018dkp, LIGOScientific:2019fpa, LIGOScientific:2020tif}.

A generic quasi-circular \bbh{} is described by 8 intrinsic parameters; the mass ratio $q=\frac{m_1}{m_2}$ (where $m_i$ are the masses of the black holes and we choose $m_1>m_2$), the spins of the individual black holes $\mathbf{S_1}$ and $\mathbf{S_2}$ and the total mass of the system $M = m_1 + m_2$ which provides the overall frequency scaling. These are the parameters of the two initial \bh{s} which coalesce to form a single spinning \bh{}, which ``rings down'' emitting \gw{s} at characteristic frequencies known as \qnm{s}. This final \bh{} can be described by its mass $M_\text{f}$ and spin $\mathbf{S_\text{f}}$. This final \bh{} is a perturbed Kerr \bh{} and the complex frequencies of the emitted \qnm{s} can be found from perturbation theory, although their relative amplitudes cannot.

Perturbation theory gives two possible values for the ringdown frequency of a perturbed \bh{}; a prograde and a retrograde frequency~\cite{Leaver:1985ax}. 
For aligned-spin systems, in cases where the original black holes are non-spinning or where their total spin $\mathbf{S} = \mathbf{S_1} + \mathbf{S_2}$ is aligned with the orbital angular momentum of the binary $\mathbf{L}$ prior to merger then the final perturbation will be in the same direction as the final \bh{'s} spin.
In these cases, the prograde frequency will therefore always dominate.
In cases where the spin $\mathbf{S}$ is anti-aligned with $\mathbf{L}$ then the final spin can be either aligned or anti-aligned with the axis of the final perturbation.
Consequently, either the prograde or the retrograde frequencies can be dominant depending on the direction of the final spin with respect to the axis of the final perturbation~\cite{Leaver:1985ax}. In the case of precessing systems, however, it is not trivial to determine which excitation will dominate the ringdown signal. 

In this paper we will provide an understanding of the interplay between the prograde and retrograde frequencies in the ringdown regime of precessing systems.
We also provide a justification for treating a precessing binary as if it has a single ringdown frequency and present an analytic formula to calculate this frequency in a frame which precesses along with the binary (known as the \emph{co-precessing frame}~\cite{Schmidt:2010it, OShaughnessy:2011pmr, Boyle:2011gg}).

There are two main issues with applying the results obtained from perturbation theory directly to \gw{} models of precessing systems.
Firstly, for a precessing binary we expect that both the prograde and retrograde \qnm{s} will contribute. However, how they interact and the region of parameter space in which each dominates is unknown. 
We will address these issues in Sec.~\ref{sec: J frame spheroidal}.
A crucial result of this work is that, although we do see oscillations in the ringdown frequency due to the superposition of two \qnm{s}, the ringdown behaviour is dominated by the average value of these oscillations.
In the event that one wishes to approximate the ringdown frequency by a single value (such as for waveform modelling purposes) we provide a simple prescription to identify the dominant frequency.
Further, we show that given the relative amplitudes of the prograde and retrograde contributions, we can make a reasonable approximate model of the oscillatory behaviour of the ringdown frequency. This can be produced using Eq.~(\ref{eqn: angular freq J}).

The second issue is that  
while the results from perturbation theory are derived in a frame in which the total angular momentum of the system is in the $z$-direction, models of \gw{s} from precessing systems often employ a non-inertial co-precessing frame in which the signal is greatly simplified. 
It is therefore important to know how the results derived using perturbation theory, namely the ringdown frequency of the system, transform when we go into the co-precessing frame.
The main result presented here is a relationship between the \qnm{} frequencies in these two frames, which we give in Eq.~(\ref{eqn: HOM expression}).
This formula will be of great assistance in modelling the \gw{s} from precessing binaries since we can now calculate the ringdown frequencies in the co-precessing frame based on a knowledge of the binary configuration without any additional modelling. 

The paper is organised as follows. 
In Sec.~\ref{sec: prelims} we discuss the conventions employed throughout the paper and recap background material regarding \qnm{s} and the co-precessing frame.
In Sec.~\ref{sec: NR} we give an overview of the numerical waveforms employed in the validation of the results presented in the rest of the paper.
In Sec.~\ref{sec: J frame} we discuss the $\mathbf{J}$-frame behaviour of the ringdown frequency and finally in Sec.~\ref{sec: cp frame} we extend this to a discussion of the behaviour in the co-precessing frame.

\section{Preliminaries}\label{sec: prelims}

In this section we will discuss the various conventions, tools and techniques required for the analysis presented in the paper. First, we discuss the different bases in which one can decompose a gravitational wave signal. We then summarise the gravitational waves emitted by a perturbed black hole. Finally we discuss the advantages of treating the signal emitted by a precessing binary in a co-precessing frame and define the co-precessing frame we will employ in the remainder of the paper.

We will work with the Newman-Penrose scalar $\Psi_4$~\cite{NP66}, which is related to the metric perturbation. This is the quantity extracted from 
the numerical simulations used to demonstrate the validity of the results in this paper, so we express everything in terms of $\Psi_4$ as a matter 
of convenience. However, these results are equally applicable to the gravitational wave strain $h$, which corresponds to a measured gravitational 
wave signal. $h$ can be obtained asymptotically from $\Psi_4$ via a double time integral.

\subsection{Decompositions}

We can decompose the \gw{} strain into multipole moments, using a variety of different bases of spin $-2$ weighted functions. The most convenient basis to use during the inspiral part of a \bbh{} coalescence is the spherical harmonic basis. During the ringdown part of the signal, one should instead use the spheroidal harmonic basis~\cite{Leaver:1985ax}.
However, it is common to decompose the complete \imr{} signal in a spherical harmonic basis when studying the complete \gw{} signal. This leads to a phenomenon known as \emph{mode mixing} in the ringdown regime~\cite{Berti:2014fga, Kelly:2012nd, London:2020uva}.

The radiative Weyl scalar $\Psi_4$ can be given as a function of time $t$ in the spherical harmonic basis by
\begin{align} \label{eqn: spherical decomp}
   \Psi_4\left(t,r,\theta,\phi\right) = {}& \frac{1}{r} \sum_{\ell, m} 
   \psi_{4, \ell m}(t) \,_{-2}Y_{\ell m}\left(\theta,\phi\right),
\end{align}
and in the spheroidal harmonic basis by
\begin{align} \label{eqn: spheroidal decomp}
   \Psi_4\left(t,r,\theta,\phi\right) = {}& \frac{1}{r} \sum_{\ell, m, n} 
   \psi^\text{S}_{4, \ell m n}(t) \,_{-2}S_{\ell m}\left(\theta,\phi,a\tilde{\omega}_{\ell m n}\right),
\end{align}
where $_{-2}Y_{\ell m}$ and $_{-2}S_{\ell m n}$ are the spin $-2$-weighted spherical and spheroidal harmonics respectively and $\left(r,\theta,\phi\right)$ are the usual spherical co-ordinates. $a\tilde{\omega}_{\ell m n}$, which appears in Eq.~(\ref{eqn: spheroidal decomp}), is the oblateness parameter. The spherical multipoles $\psi_{4, \ell m}(t)$ can be obtained directly from the orthogonality of the spherical harmonics. However, at the present time, the spheroidal multipoles $\psi^\text{S}_{4, \ell m n}(t)$ cannot and must instead be estimated from the spherical multipoles themselves or via other means, such as time domain fitting~\cite{London:2020uva,London:2014cma,Li:2021wgz}. 

The spheroidal harmonics can be written as a linear combination of the spherical harmonics as
\begin{align} 
   _{-2}S_{\ell m n}\left(\theta,\phi\right) = {}& 
   \sum_{\ell'} \alpha_{\ell \ell' m} \,_{-2}Y^*_{\ell m}\left(\theta,\phi\right)
\end{align}
where $\alpha_{\ell \ell' m}$ are the mixing co-efficients which measure the overlap between a given set of spherical and spheroidal harmonics~\cite{Berti:2014fga,London:2014cma}. From this we can get an expression for the spherical harmonic multipoles in terms of the spheroidal harmonic multipoles and vice versa.
We can therefore study the ringdown signal as a set of spheroidal multipole moments where we do not need to consider the effect of mode mixing and use these to construct a consistent \imr{} description of the \gw{} signal in terms of the spherical harmonic multipoles only.

We will first consider the waveform in terms of the spheroidal harmonics (in Sec.~\ref{sec: J frame spheroidal}) where our results will be cleanest since we are considering a ringdown effect.
We will then consider the effect of mode mixing when extending to the spherical harmonic basis in Sec.~\ref{sec: spherical mode mixing}.
The results presented in Sec.~\ref{sec: cp frame} are applicable in either basis but the comparison to numerical data is made in terms of the spherical harmonic basis.

\subsection{Formation of the final black hole}\label{sec: final BH formation}

The frequencies of the \qnm{s} emitted by a perturbed \bh{} are generally determined using perturbation theory in the frame in which the spin of the black hole is taken to be in the $z$-direction: $\mathbf{S_\text{f}} \parallel \hat{z}$. 
For a non-spinning or aligned-spin binary, this corresponds to an inertial frame in which the black holes orbit in the $x-y$ plane prior to merger. 
In the case of a precessing binary the orbital plane precesses and relating the final spin direction to the inclination of the orbital plane just before merger is not trivial.

The \qnm{s} are characterised by $\ell$, $m$ and $n$. In the spheroidal basis, a given multipole moment $\psi^\mathrm{S}_{4, \ell m n}$ can be written as the sum of all the \qnm{s} with the same values of $\ell$, $m$ and $n$\footnote{In fact, a given multipole moment can be written as the sum over all \qnm{s} with the same value of $m$. However, the contribution from terms with $\ell' \neq \ell$ and $n' \neq n$ are strongly suppressed~\cite{London:2020uva}.}.
In this work we will consider only contributions from the fundamental \qnm{s} and neglect contributions from \qnm{s} with $n>0$.
As discussed in the previous subsection, a given spherical multipole moment $\psi_{4,\ell m}$ contains contributions from all spheroidal multipoles with the same value of $m$. 
Consequently, the following discussion is only strictly true when considering the signal decomposed in the spheroidal basis. 

There are two \qnm{} frequencies associated with each set of $\ell$, $m$ and $n$ --- the prograde \qnm{s} and the retrograde \qnm{s}. The prograde modes are excited if the perturbation on the Kerr \bh{} is in the same direction as the spin. Conversely, the retrograde modes are excited if the perturbation is in the opposite direction to the spin.
For a remnant \bh{} formed by a \bbh{} merger, we assume that the perturbation will be in the direction in which the original \bh{s} were orbiting.

For aligned-spin systems, one traditionally makes the approximation that either the prograde or retrograde \qnm{s} are excited, but not both; see, 
for example, Ref.~\cite{Husa:2015iqa}. 
In principle, both could be excited but this depends on details of the merger that are not fully understood.
There is some recent evidence for both prograde and retrograde modes being excited, though one dominates strongly over the 
other~\cite{Li:2021wgz,Bernuzzi:2010ty,Barausse:2011kb,Dhani:2020nik,Ma:2022wpv}.
As in Ref.~\cite{Ma:2022wpv}, we note that Ref.~\cite{Dhani:2020nik} finds evidence of retrograde modes where we do not (most notably for equal mass, non-spinning systems).
Nevertheless, in the majority of the aligned-spin parameter space, we consider it a reasonable approximation that only one frequency is meaningfully excited and more generally true that one will clearly dominate over the other. 

Which is dominant depends on the direction of the final spin $\mathbf{S}_\mathrm{f}$ with respect to the direction in which the \bh{s} were originally orbiting, as characterised by $\mathbf{L}$. 
This determines the direction of the perturbation with respect to the final spin direction.
If the remnant \bh{} is spinning in the same direction as the perturbation (so the final spin direction is aligned with the axis of the perturbation) then we say the \bh{} has ``positive'' final spin.
Similarly, if the final spin is anti-aligned with the axis of the perturbation then we say it has ``negative'' final spin.
For aligned-spin binaries, this is equivalent to to the final spin being aligned or anti-aligned with $\mathbf{L}$.
This can be predicted using fits to numerical data; see, for example Ref.~\cite{Varma:2018aht} and references therein.
The transition from cases where the dominant ringdown frequency is prograde to retrograde will occur when the final spin of the \bh{} is zero.

For precessing systems, the picture is somewhat more complicated; the final spin need not be either aligned nor anti-aligned with respect to $\mathbf{L}$ nor with respect to the axis of the perturbation and can instead be arbitrarily oriented.
Consequently, both the prograde and retrograde modes will be excited to some extent. 
In some cases, they may be excited to approximately equal amplitudes.
Drawing from the non-precessing case, we would expect the prograde modes to dominate for cases with a ``positive'' final spin and for the retrograde modes to dominate for the cases with ``negative'' final spin.
The interpretation of a ``positive'' or ``negative'' final spin, along with its implication for the region of parameter space in which we observe either the prograde or the retrograde modes to dominate is discussed at length in Sec.~\ref{sec: J frame spheroidal}.
We also study the effect of the excitation of \emph{both} prograde and retrograde perturbations on the ringdown signal and provide a simple model which captures the most important features of these effects for use in signal modelling.

\subsection{Co-precessing frame}

We now consider the complete \imr{} gravitational wave signal from a precessing binary and the advantages of using the co-precessing frame. Since we are considering the complete signal, this discussion applies to the spherical harmonic multipoles.

So far we have considered the system in an inertial frame in which $\mathbf{S}_\mathrm{f} \parallel \hat{z}$.
Taking the direction of the total angular momentum $\mathbf{J}$ of the binary to be fixed throughout the binary's evolution,
this corresponds to an inertial frame in which $\mathbf{J} = \mathbf{L} + \mathbf{S_1} + \mathbf{S_2}$ (where $\mathbf{L}$ is the orbital angular momentum of the binary) is along $\hat{z}$ throughout the binary's evolution\footnote{For most precessing systems, the direction of $\mathbf{J}$ will remain approximately constant throughout the evolution of the binary. This is known as simple precession~\cite{Apostolatos:1994mx,Kidder:1995zr}.}.
We therefore refer to this frame as the $\mathbf{J}$-frame.

An inertial \textbf{J}-frame is a reasonable frame in which to model the \gw{} signal emitted during the inspiral phase by non-spinning or aligned-spin binaries. 
However, for precessing binaries it is not an ideal frame in which to model the \gw{s}; the signal is very complicated and shows oscillations in the amplitude and phase~\cite{Apostolatos:1994mx,Kidder:1995zr}. 
Instead, it is more convenient to model the signal from precessing binaries in a non-inertial \emph{co-precessing} frame~\cite{Hannam:2013oca, Khan:2018fmp, Khan:2019kot, Pratten:2020ceb, Hamilton:2021pkf, Pan:2013rra, Taracchini:2013rva, Ossokine:2020kjp, Estelles:2021gvs,Blackman:2017pcm,Blackman:2017pcm}, which tracks the precession of the binary and where the waveform resembles that of a non-precessing system~\cite{Schmidt:2010it}. The waveform is then rotated back into the $\mathbf{J}$-frame in order to produce the final model.

There are a number of different ways one can choose to define the co-precessing frame; such as tracking the direction of the orbital angular momentum (calculated from Newtonian or \pn{} estimates of varying order) or tracking the optimal emission direction.
While these directions are very similar, they are not identical~\cite{Hamilton:2018fxk}. In the rest of the paper we consider only the co-precessing frame that tracks the optimal emission direction. In the case where the optimal emission direction is calculated using only the $\ell=2$ multipoles this frame is also known as the quadrupole aligned frame.

The spherical harmonic multipoles transform under rotations by the Euler angles $\{\alpha,\beta,\gamma\}$ as~\cite{Bruegmann:2006ulg, WignerEugenePaul1959Gtai} 
\begin{align}\label{eqn: multipole rotation}
   \psi'_{4,\ell m'} = {}& \sum_{m=-\ell}^\ell e^{im\alpha} d^\ell _{m' m}\left(-\beta\right) e^{im'\gamma} \psi_{4,\ell m},
\end{align}
where $\psi_{4,\ell m}$ are the multipoles in the initial frame and $\psi'_{4,\ell m}$ are the multipoles in the new frame. $d^\ell _{m' m}$ are the Wigner $d$-matrices. If we take the initial frame to be the frame in which the total angular momentum of the system is aligned with the $z$-direction and the new frame to be the co-precessing frame which tracks the direction of maximum emission, then $\{\alpha,\beta,\gamma\}$ are the set of precession angles which can be found as described in~\cite{Schmidt:2010it, OShaughnessy:2011pmr, Boyle:2011gg}. In summary,
\begin{align} 
   \tan\alpha = {}& \frac{V_x}{V_y}, \\
   \cos\beta = {}& \frac{V_z}{|\mathbf{V}|}, \\
   \dot\gamma = {}& -\dot\alpha \cos\beta, \label{eqn: mrc}
\end{align}
where $\mathbf{\hat{V}}$ is the optimal emission direction and Eq.~(\ref{eqn: mrc}) is known as the minimal rotation condition. 
In Sec.~\ref{sec: cp frame}, where we have imposed a convention on $\beta$, the minimal rotation condition is more appropriately written with $|\cos\beta|$.

In this paper we approximate the ringdown behaviour of $\alpha$ and $\beta$ as a straight line and a constant respectively. The value of $\alpha$ can be determined entirely from the \qnm{} frequencies while $\beta$ can at present only be found numerically. The validity of these approximations are examined in the following paragraphs. 

Starting from Eq.~(\ref{eqn: multipole rotation}) and assuming that we have only the $(2,2)$ multipole in the co-precessing frame then, for sufficiently small $\beta$, we have the following relation between the spherical harmonic multipoles in the $\mathbf{J}$-frame and the co-precessing frame,
\begin{align} \label{eqn: dominant multipole transformation}
   \psi^\text{J}_{4,2m} = {}& e^{-i\left(m\alpha+2\gamma\right)} \, d^2_{2 m} \, \psi^\text{cp}_{4,22}, 
\end{align}
from which we can derive approximate expressions for $\alpha$ and $\beta$ in terms of the amplitudes $A_{\ell m}$ and phases $\phi_{\ell m}$ 
of the spherical harmonic multipoles in the $\mathbf{J}$-frame~\cite{OShaughnessy:2012iol,Marsat:2018oam,Ossokine:2020kjp,Estelles:2020osj,Hamilton:2021pkf}. These are
\begin{align} 
   \alpha = {}& \phi_{22} - \phi_{21}, \label{eqn: alpha approx} \\
   \tan\frac{\beta}{2} = {}& \frac{A_{21}}{2A_{22}}. \label{eqn: beta approx}
\end{align}

\begin{figure}[t]
   \centering
   \includegraphics[width=0.48\textwidth]{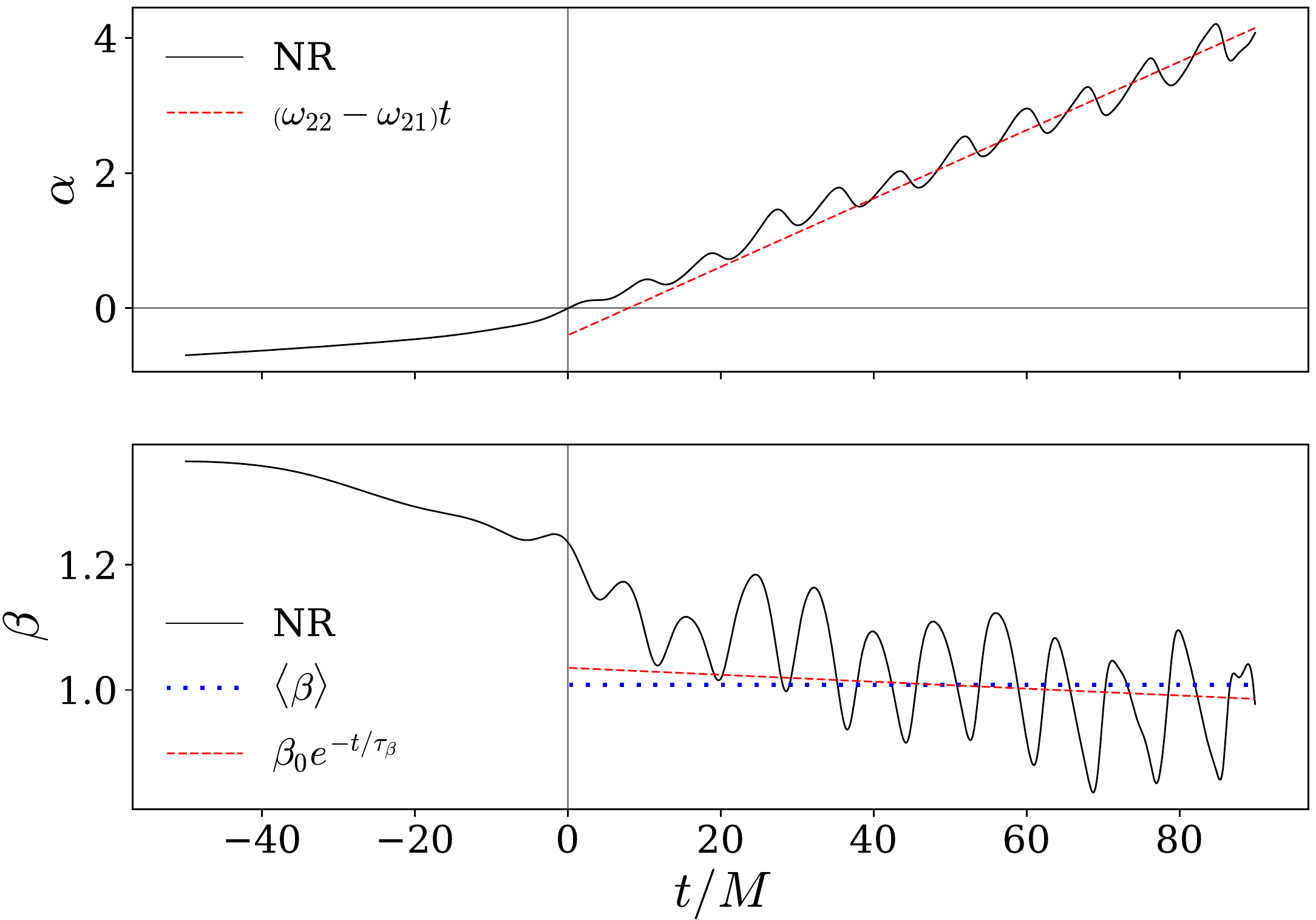} 
   \caption{A comparison of the precession angles $\alpha$ and $\beta$ calculated from the NR $\ell=2$ multipoles with those given by the approximations in Eqs.~(\ref{eqn: beta approx}) and~(\ref{eqn: alphadot}) respectively for the case CF21-79 $\left(q,\chi,\theta_\text{LS}\right) = \left(8,0.8,120^\circ\right)$. The grey vertical line indicates merger, defined as the maximum of the sum of the square of the amplitude of the $\ell=2$ multipoles. The red dashed lines give the predicted value of $\alpha$ and $\beta$ based on Eqs.~\ref{eqn: alphadot} and~\ref{eqn: beta approx} respectively. The blue dotted line gives the mean value of $\beta$.}
   \label{fig: RD precession angles}
\end{figure}

In the ringdown we can write the $n=0$ multipoles in the spheroidal harmonic basis as 
\begin{align}
   \psi_{4,\ell m 0} = {}& A_{\ell m 0} e^{-i\phi_{\ell m 0}}.
\end{align}
Since in this paper we are considering only the $n=0$ multipoles, from here on we will drop the labelling of the multipoles by $n$. 
The multipoles can then be written
\begin{align}
   \psi_{4,\ell m} = {}& A_{\ell m} e^{-i\phi_{\ell m}}
   = \mathcal{A}_{\ell m}e^{-\frac{t}{\tau_{\ell m}}}e^{-i\omega_{\ell m}t}, \label{eq:RD}
\end{align}
where $\mathcal{A}_{\ell m}$ are the time-independent amplitudes of the \qnm{s}, $\tau_{\ell m}$ are the damping times of the \qnm{s} and $\omega_{\ell m}$ are the \qnm{} frequencies. We do not expect  mixing to have a strong effect for either the $(2,|2|)$ or the $(2,|1|)$ multipoles so can assume the spherical harmonic multipoles are also described by the above expression. We can therefore see that under the assumption that we have only the $(2,|2|)$ multipoles in the co-precessing frame, the ringdown precession angle $\alpha$ is linear with gradient 
\begin{align}\label{eqn: alphadot}
   \dot\alpha = {}& \omega_{22} - \omega_{21}.
\end{align}
We can similarly see that $\beta$ decays exponentially with a damping time 
\begin{align} \label{eqn: taubeta}
   \tau_\beta = \frac{\tau_{22}\tau_{21}}{\tau_{21}-\tau_{22}}.
\end{align}
Thus if $\tau_{22} \approx \tau_{21}$, which is generally the case, then the value of $\beta$ will be approximately constant throughout the ringdown.

It should be noted that in the derivation of the above expressions (Eqs.~\ref{eqn: dominant multipole transformation}--\ref{eqn: taubeta}) we have assumed that we have only the (2,2) multipole in the co-precessing frame. 
Due to the independence of the rotation of multipoles with a given $\ell$, we can equivalently make the assumption that for a given set of $\ell$-multipoles we have only the $(\ell,\ell)$ multipole in the co-precessing frame. 
We would then find that instead of Eq.~\ref{eqn: alphadot}, we find
 \begin{align}
    \dot\alpha = {}& \omega_{\ell\ell} - \omega_{\ell,\ell-1}.
\end{align}
In what follows, when an approximation for $\dot\alpha$ is required we will use Eq.~\ref{eqn: alphadot} for simplicity.

\begin{figure*}[htbp]
   \centering
   \includegraphics[width=\textwidth]{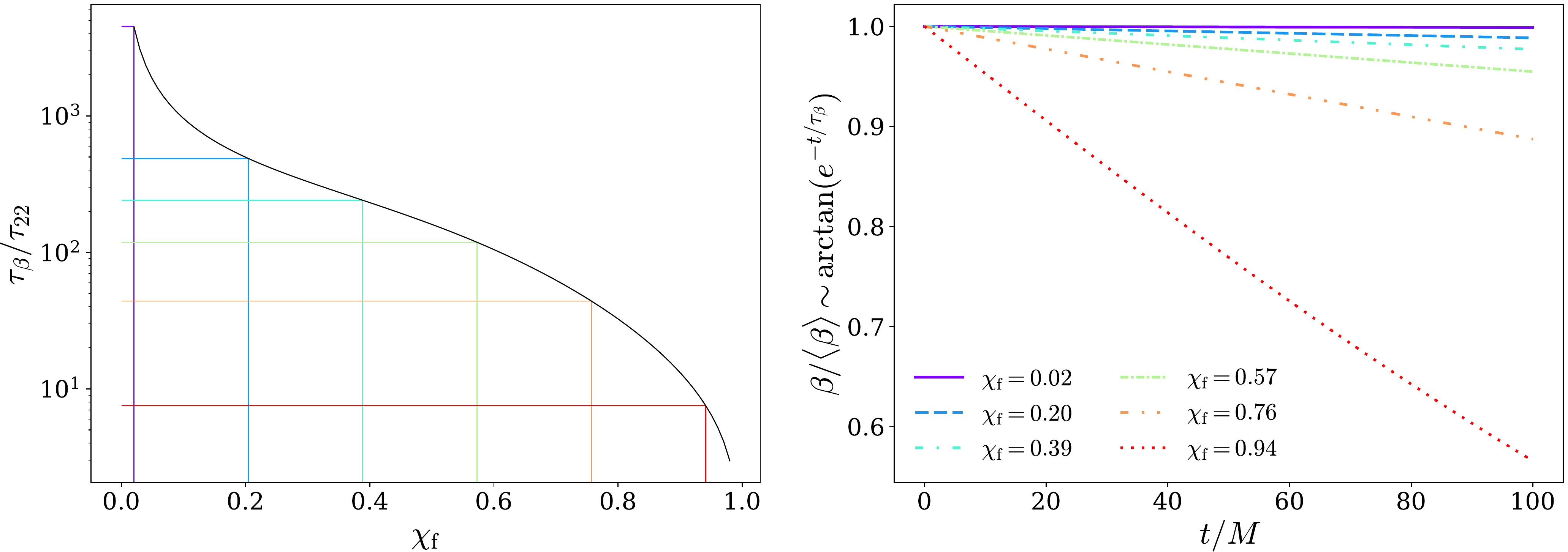} 
   \caption{The left hand panel shows the ratio of the rate of decay of $\beta$ to that of the (2,2) multipole, $\frac{\tau_\beta}{\tau_{22}}$, post-merger as a function of the spin of the final black hole $\chi_\mathrm{f}$. The right hand panel gives a sense of the slope of $\beta$ for a given set of final spins. The form of $\tau_\beta$ is given by Eq.~(\ref{eqn: taubeta}). In all cases we assume a ``positive'' final spin.
   }
   \label{fig: tau beta}
\end{figure*}

The validity of these approximations is demonstrated in Fig.~\ref{fig: RD precession angles}, where we have compared our approximations with the precession angles calculated using the $\ell=2$ multipoles of a \nr{} waveform. We can see in the upper panel that, neglecting the oscillations, the gradient of $\alpha$ is indeed constant through ringdown and is fairly well captured by Eq.~(\ref{eqn: alphadot}). In the lower panel we can see that while the exponential decay of $\beta$ due to the \qnm{} damping times may capture the trend of the ringdown angle marginally better, it is reasonable to neglect it and treat the ringdown value of $\beta$ as a constant to within numerical error. 

Fig.~\ref{fig: tau beta} shows how the rate of decay of $\beta$ through ringdown varies with the dimensionless spin of the final black hole $\chi_\text{f}$. As can be seen from the left hand panel, the rate at which $\beta$ decays is a tiny fraction of the rate at which the multipoles themselves decay. It is therefore reasonable to treat the ringdown value of $\beta$ as constant. From the right hand panel we can indeed see that for spins up to $\chi_\text{f} \sim 0.75$, $\beta$ loses less than 10\% of its value over 100$M$ following merger, further validating this assumption. It should be noted that for large final spins, $\beta$ loses an appreciable fraction of its value over this period (for $\chi_\text{f}=0.94$ $\beta$ loses around 40\% of its value) . However, it is very unusual to have systems with both a large value of $\beta$ and such a high final spin value~\cite{BAM-catalog, Mroue:2013xna, Boyle:2019kee, Healy:2017psd, Healy:2019jyf, Healy:2020vre, Healy:2022wdn, Jani:2016wkt}. 
Furthermore, as can be seen from the left hand panel, even for such high spins, the decay rate of the (2,2) multipole is still around 10 times faster than that for $\beta$ so the multipoles will have decayed to around $0.02\%$ of their original value $100M$ after merger.
Consequently, the absolute change in magnitude of $\beta$ for systems with large $\chi_\text{f}$ is generally very small and we can still make the approximation that $\beta$ is constant.

\section{NR waveforms}\label{sec: NR}

The numerical waveforms used in the paper were produced using the BAM code~\cite{Bruegmann:2006ulg}.
A subset of these waveforms are taken from a catalogue of 80 precessing BAM waveforms~\cite{BAM-catalog}. This catalogue consists of single-spin systems where the spin is placed on the larger black hole. It covers the parameter space up to mass ratio $q=8$, dimensionless spin magnitude $\chi = |\mathbf{S_1}|/m_1^2 = 0.8$ and with five equally spaced values of spin orientation $\theta_\text{LS}$, measured as the angle between the total spin and the orbital angular momentum. 
Further details can be found in Ref.~\cite{BAM-catalog}.

The region of parameter space covered by this, and indeed any, \nr{} catalogue in which most of the effects discussed in this paper are sufficiently large to be noticeable is fairly restricted. We therefore focus on a set of cases where these effects are most pronounced. 
This is the region with higher mass ratio, larger spin magnitude and greater spin opening angle. 
To supplement the waveforms taken from the catalogue, we produced four additional waveforms in the region of parameter space of greatest interest.
These waveforms were produced following the same procedure as described in Ref.~\cite{BAM-catalog}. 
A summary of the properties of the cases studied in detail in this paper is given in Tab.~\ref{tab: NR cases}.

We work with the time domain $\Psi_4$ data, which is calculated directly using the \nr{} evolution. This data is decomposed into spherical multipoles as described by Eq.~(\ref{eqn: spherical decomp}). We remove both the junk radiation at the start of the waveform (due to imperfect initial data) and the post-ringdown waveform where the exponential decay has fallen below the noise floor (generally around $100M$ after merger) using a standard Hann window function. The data are then resampled to have a uniform time step of 0.15M.

\begin{table}[t]
   \centering
   \begin{tabular}{@{} lccccc @{}} 
      \toprule
      Simulation ID & $q$ & $\chi$ & $\theta_\text{LS} (^\circ)$ & $M_\text{f}$ & $\chi_f$ \\
      \colrule
      --- & 1& 0.0 & --- & 0.952 & 0.686 \\
      \colrule
      CF21-6 & 1 & 0.4 & 30 & 0.946 & 0.740 \\
      \colrule
      --- & 8 & 0.8 & 110 & 0.989 & 0.595 \\
      CF21-79 & 8 & 0.8 & 120 & 0.988 & 0.548 \\
      --- & 8 & 0.8 & 130 & 0.991 & 0.480 \\
      --- & 8 & 0.8 & 140 & 0.991 & 0.416 \\
      CF21-80 & 8 & 0.8 & 150 & 0.991 & 0.371 \\
      --- & 8 & 0.8 & 160 & 0.992 & 0.315 \\
      \botrule
   \end{tabular}
   \caption{Summary of the properties of the key \nr{} cases focussed on in this paper. For details of other cases see Ref~\cite{BAM-catalog}.}
   \label{tab: NR cases}
\end{table}

The configurations are each parameterised by the masses and spins of the initial two black holes in the binary and by the mass and spin of the final black hole. The properties of the binary are defined as part of the initial data of the simulation, while the final mass and spin are taken from their numerical values at the end of the simulation.

We consider our data in two different frames; the co-precessing frame, and an approximately inertial frame, which we refer to as the ``$\mathbf{J}$-frame'' and define as follows. 
The direction of $\mathbf{J}$ is approximately constant throughout the evolution of the binary.
We track this direction and transform into a frame in which $\mathbf{J}$ is along the $z$-axis at all times.
In this frame the final spin of the black hole $\mathbf{S}_\text{f} = \mathbf{J}_\text{f}$ is aligned with the $z$-direction.
We calculate the final-$\mathbf{J}$ direction to be $\mathbf{J}_\text{f} = \mathbf{J}_\text{i} - \mathbf{J}_\text{rad}$, where $\mathbf{J}_\text{i}$ is the initial value of the total angular momentum and $\mathbf{J}_\text{rad}$ is the angular momentum radiated throughout the evolution of the binary.

We define the co-precessing frame to be that which tracks the optimal emission direction. We calculate the precession angles $\{\alpha,\beta,\gamma\}$ that describe the rotation from the $\mathbf{J}$-frame to the co-precessing frame using the method presented in~\cite{OShaughnessy:2011pmr}. Unless otherwise stated, we use all the spherical multipoles up to $\ell=4$ to calculate the optimum emission direction and thus prescribe the co-precessing frame. 
This enables us to examine all multipoles up to $\ell=4$ in the co-precessing frame.
However, when calculating a mean value of the ringdown $\beta$ from the \nr{} data, we use only the $\ell=2$ multipoles to calculate the optimum emission direction. This is consistent with the approximations made when finding an expression for $\beta$ in terms of the multipoles in Sec.~\ref{sec: prelims}. 
In the time domain, the co-precessing frames defined by using the $\ell=2$ and $\ell \le 4$ multipoles are very similar.

We define merger to be the point in the waveform at which the quantity 
\begin{align}
   \overline{A} = {}& \sum_{m=-2}^2 |A_{2,m}\left(t\right)|^2
\end{align}
is maximised, where $A_{2,m}$ are the amplitudes of the $\ell=2$ spherical multipoles. We then set $t=0$ at merger.

We calculate the angular frequency of each of the post-merger spherical multipoles in both the $\mathbf{J}$-frame and the co-precessing frame. This is given by the time derivative of the phase of each of the respective multipoles:
\begin{align} 
   \omega^F_{\ell m} = {}& \frac{\text{d}}{\text{d}t} \phi^F_{\ell m},
\end{align}
where $F$ indicates the frame in which we are considering the data.

The amplitude of each of the \qnm{s} contained in a given spherical multipole in the $\mathbf{J}$-frame can be estimated by fitting a series of damped sinusoids to the post-merger data over the approximate region from $t=10M$ to $t=100M$. These amplitudes are the product of the intrinsic amplitude of the \qnm{} and the relevant mixing co-efficient. 
More details on the method used are given in~\cite{London:2014cma}, where the amplitude of the \qnm{s} are fit while the \qnm{} frequencies are fixed according to Leaver's method as described in~\cite{Leaver:1985ax}.

\section{$\mathbf{J}$-frame behaviour}\label{sec: J frame}

Having established the various frames in which we plan to consider the ringdown signal, we now examine the ringdown frequency in each of these frames. We first consider the signal in the $\mathbf{J}$-frame, since this is the frame in which the results from perturbation theory are most applicable. In Sec.~\ref{sec: cp frame} we will then consider the effect of rotations into the co-precessing frame.

Further, as discussed in Sec.~\ref{sec: prelims}, the \gw{s} emitted by a \bh{} which is ringing down can be decomposed into either a spheroidal or a spherical harmonic basis. The most natural basis in which to describe the \gw{s} from a single perturbed \bh{} is the spheroidal harmonic basis, so we consider this decomposition first.
However, as previously discussed, it is often more convenient to consider the spherical harmonic decomposition.
We therefore also discuss the effect of mode mixing due to this choice of basis on the results presented here.

\subsection{Spheroidal harmonic picture}\label{sec: J frame spheroidal}

When decomposing the \gw{s} emitted by a perturbed Kerr \bh{} in a spheroidal harmonic basis, each of the $\left(\ell,m\right)$-multipoles are a superposition of the relevant \qnm{s} (i.e. those with the same $\ell$ and $|m|$ values). If we consider just the dominant \qnm{s} (those with $n=0$) then the $\left(\ell,m\right)$-multipoles are given by the sum of the prograde and retrograde \qnm{s}. We therefore have~\cite{Berti:2005ys, Lim:2022veo, Dhani:2020nik}
\begin{align}\label{eqn: pro retro sum}
   \psi_{4,\ell m} = {}& A^{\ell m}_+ e^{-i\phi^{\ell m}_+} + A^{\ell m}_- e^{-i\phi^{\ell m}_-},
\end{align}
where $A^{\ell m}_\pm$ and $\phi^{\ell m}_\pm$ are the amplitudes and phases of the prograde and retrograde \qnm{s} respectively. We have dropped the subscript $n$ for simplicity. The phases are given by $\phi^{\ell m}_\pm = \omega^{\ell m}_\pm t$, where $\omega^{\ell m}_\pm$ are the 
 \qnm{} angular frequencies and we are using the convention $\omega^{\ell m}_+ > 0$ and $\omega^{\ell m}_- < 0$. The amplitudes $A^{\ell m}_\pm$ cannot be found using perturbation theory, only the \qnm{} frequencies and damping times, which are absorbed into $A_\pm$, as in 
 Eq.~(\ref{eq:RD}). 

\begin{figure}[t]
   \centering
   \includegraphics[width=0.48\textwidth]{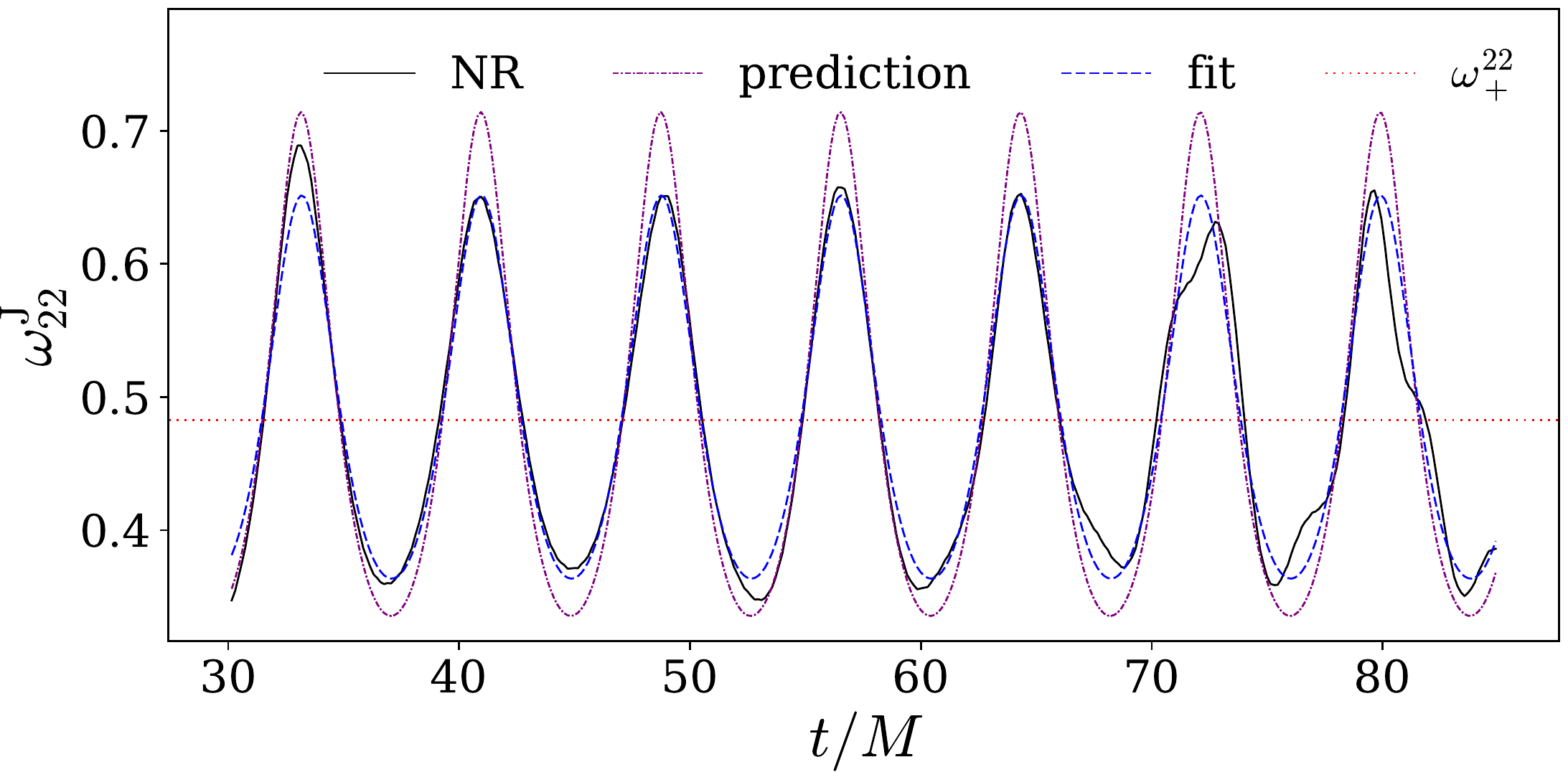} \\
   \includegraphics[width=0.48\textwidth]{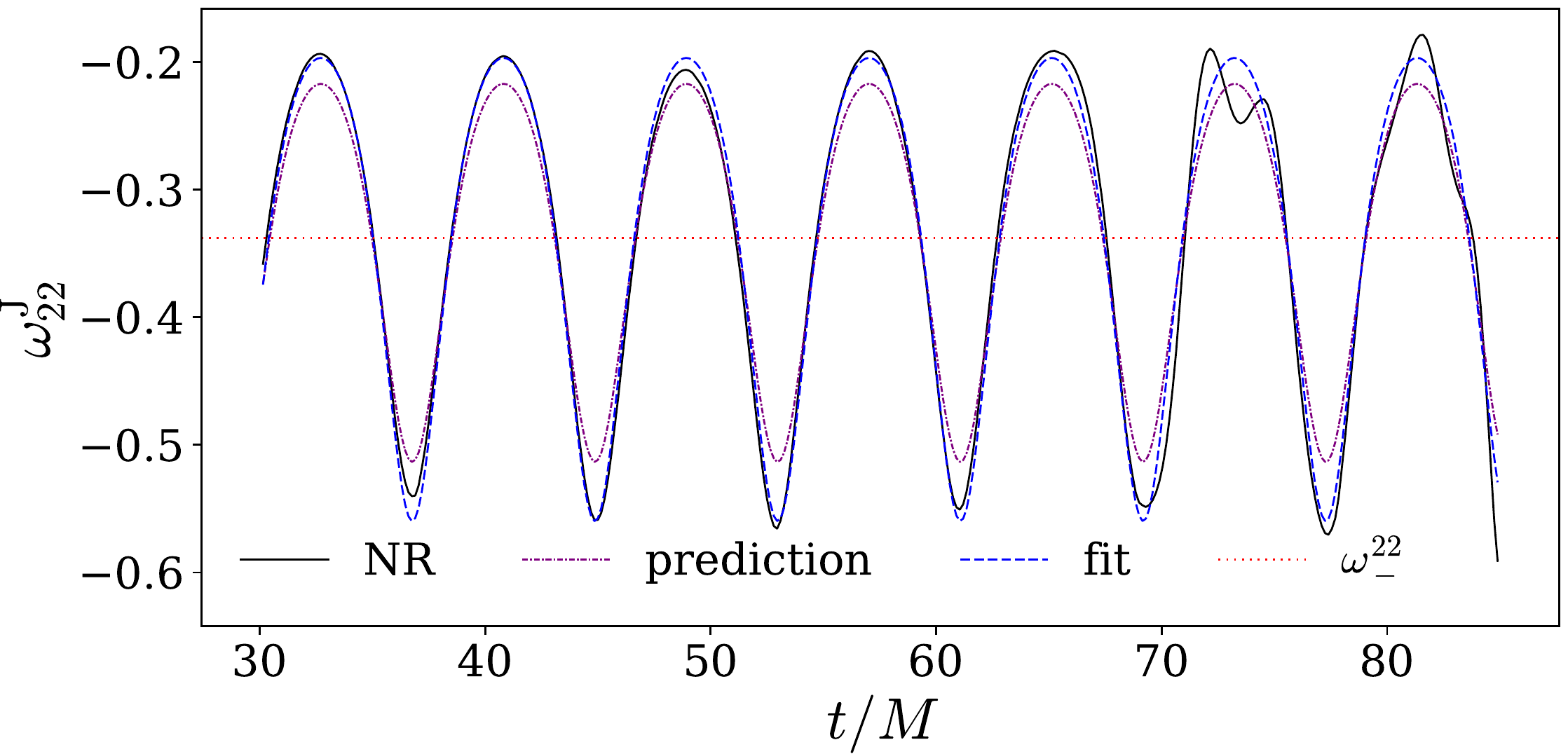}
   \caption{A comparison of the time domain angular frequency of the (2,2) multipole in the $\mathbf{J}$-frame with that predicted by Eq.~(\ref{eqn: angular freq J}). The top panel shows the case CF21-79, the bottom panel shows CF21-80. The prediction shown in the dot-dashed purple lines uses the relative contribution from the prograde and retrograde frequencies found by fitting a series of damped sinusoids to the individual spherical multipoles. This gives a ratio of $A_+ : A_-$ of $4.49:1$ and $0.185:1$ respectively for the two cases respectively. The dashed blue line shows the result of fitting Eq.~(\ref{eqn: angular freq J}) to the data. In this case we find values of $A_+ : A_-$ of $5.78:1$ and $0.222:1$. The red line gives the frequency of the dominant \qnm{}. In all cases, the \qnm{} frequencies are calculated using perturbation theory.
   }
   \label{fig: oscillatory angular frequency}
\end{figure}

This superposition of the prograde and retrograde \qnm{s} manifests as oscillations in the phase and frequency of each multipole about a mean value. This is most easily seen in the expression for the angular frequency of a given multipole.
The phase of each multipole $\phi_{\ell m}$ is given by 
\begin{align}\label{eqn: J frame multipole phase}
   \tan\phi_{\ell m} = {}& \frac{A^{\ell m}_+\sin\phi^{\ell m}_+ + A^{\ell m}_-\sin\phi^{\ell m}_-}
   {A^{\ell m}_+\cos\phi^{\ell m}_+ + A^{\ell m}_-\cos\phi^{\ell m}_-}.
\end{align} 
We assume the damping times of the prograde and retrograde frequencies are approximately equal and thus the time dependence of $A^{\ell m}_\pm$ can be factored out. For the parameter space considered in this study, we find that the greatest dimensionless spin magnitude of the final \bh{} is $\chi_\text{f} = 0.86$, which corresponds to a maximum 25\% difference between the prograde and retrograde damping times.
For the majority of cases, $\chi_\text{f} < 0.7$, corresponding to a difference in the damping times of less than 10\%. We therefore consider it to be a reasonable approximation to factor out the time dependence of the $A^{\ell m}_\pm$. 
The validaty of this approximation is demonstrated in Fig.~\ref{fig: oscillatory angular frequency}.
Differentiating Eq.~(\ref{eqn: J frame multipole phase}) with respect to time (and dropping the $\ell m$ superscript for simplicity) we find
\begin{align}\label{eqn: angular freq J}
   \dot{\phi} = {}& \frac{A_+^2 \omega_+ + A_-^2 \omega_- + A_+ A_- \left(\omega_+ + \omega_-\right)
   \cos\left[\left(\omega_+ - \omega_-\right)t\right]}
   {A_+^2 + A_-^2 + 2 A_+ A_- \cos\left[\left(\omega_+ - \omega_-\right)t\right]}.
\end{align}
In the rest of the paper we will focus on the angular frequency of each multipole moment as method of investigating the \qnm{} content, rather than the more traditional approach of identifying the \qnm{} content in the multipole as represented in Eq.~\ref{eqn: pro retro sum}.

The mean value of the angular frequency is then
\begin{align}\label{eqn: pro retro qnm selection}
   \langle\dot\phi\rangle  = {}& 
   \begin{cases}
      \omega_+ & A_+ > A_- \\
      \omega_- & A_+ < A_- \\
      \frac{1}{2}\left(\omega_+ + \omega_-\right) & A_+ = A_-
   \end{cases}.
\end{align}

The extrema of the oscillations are given by
\begin{align} 
   \dot{\phi}_\text{ext} = {}& \frac{A_+ \omega_+ \pm A_- \omega_-}{A_+ \pm A_-}.
\end{align}

It is worth noting that the average value of the angular frequency is either exactly the prograde or the retrograde frequency, except in the special case that the two contributions are equal. 
This is why, if we just look at the mean ringdown frequency (which is common in NR analyses,
where it can be unclear whether any oscillations present in the data are physical effects or are instead due to numerical noise) we will see only a single frequency.
Similarly, when Fourier transforming the post-merger waveform to obtain the frequency-domain ringdown signal we see a single peak corresponding to the mean ringdown frequency. 
The oscillatory behaviour seen in the time-domain waveform frequency is distributed across a range of frequencies and consequently is not clearly apparent in the frequency domain waveform. 
When modelling features of frequency-domain waveforms that correspond to the ringdown frequency (e.g. the location of the dip in the phase derivative) we therefore need only consider the behaviour of this mean value.

To gain a clear understanding of the behaviour of this mean frequency we need to know where in parameter space the transition from prograde to retrograde values happens. This transition occurs when the amplitude of the prograde and retrograde frequencies are equal (i.e. $A_+ = A_-$). 
However, this information cannot be obtained analytically and the part of the parameter space in which $A_+$ and $A_-$ are of comparable magnitude is particularly sparsely sampled by \nr{} waveforms.
We therefore do not have sufficient numerical data to accurately determine the transition point in this fashion and must instead rely on an approximation.
In the non-precessing case, the transition from prograde to retrograde is expected to occur when the final spin goes from ``positive'' to ``negative'' with respect to the orbital plane of the binary.
It is therefore reasonable to assume that a similar condition will determine the transition point for precessing systems.

There is a dearth of information in the literature describing the magnitude and direction of the final spin for precessing binaries with mass ratios above $q=4$ (precisely the region we are concerned with). Final spin fits for precessing binaries do exist but they have little or no tuning to numerical simulations as far as mass ratio $q=8$~\cite{Hofmann:2016yih,Varma:2018aht,Haegel:2019uop}. 
Fits for aligned spin binaries have however been tuned to \nr{} as far as $q=18$~\cite{Husa:2015iqa,Jimenez-Forteza:2016oae}.
One reasonable approximation to obtain the magnitude and direction of the final spin for a precessing binary is then
\begin{align} \label{eqn: final spin}
   S_\mathrm{f} = 
   {}& \sqrt{\left(S_\mathrm{f}^\parallel\right)^2 + \left(S_\mathrm{f}^\perp\right)^2}, \\
   \cos\theta_\mathrm{f} = 
   {}& \cos\left(\mathbf{\hat{L}}\cdot \mathbf{S_\mathrm{f}}\right) =
   \frac{S_\mathrm{f}^\parallel}{S_\mathrm{f}},
\end{align}
where $S_\mathrm{f}^\parallel$ is given by an aligned-spin final spin fit and $S_\mathrm{f}^\perp$ is the component of the spin in the orbital plane during inspiral, which is assumed not to change throughout the evolution of the binary. 
This, or a similar\footnote{In Ref.~\cite{Pratten:2020ceb} a \pn{} estimate for the magnitude of the orbital angular momentum is included in the expression for the final spin magnitude (see Eq. (4.17)). This does not significantly alter the statements made in this section.}, assumption has been previously implemented in a number of precessing models, such as Refs.~\cite{Hannam:2013oca,Khan:2018fmp,Khan:2019kot,Pratten:2020ceb,Hamilton:2021pkf}, and has been found to be very accurate-- the value predicted by Eq.~(\ref{eqn: final spin}) agrees with the \nr{} final spin estimate to within $1.5\%$ for all 80 cases presented in Ref.~\cite{BAM-catalog}.

\begin{figure}[t]
   \centering
   \includegraphics[width=0.48\textwidth]{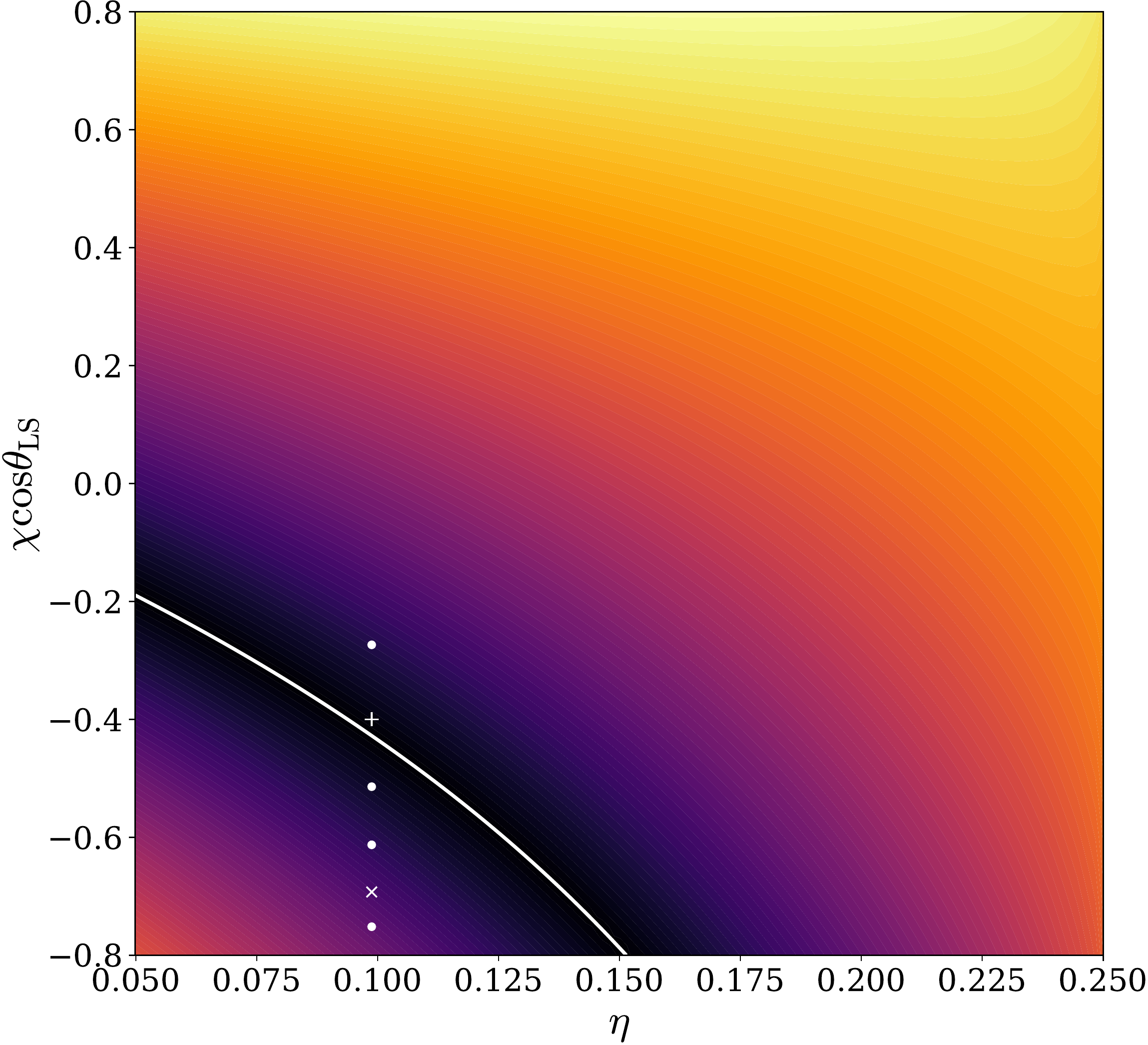} 
   \caption{The magnitude of $\mathbf{S}^\parallel_\text{f}$ as a function of the mass ratio and aligned-spin component of the initial binary. The white line indicates the line along which $\mathbf{S}^\parallel_\text{f} = 0$. The white $+$ and $\times$ indicate the position in parameter space of the CF21-79 and CF21-80 cases shown in Fig.~\ref{fig: oscillatory angular frequency} respectively. The white dots show the position of cases with $(q,\chi)=(8,0.8)$ and $\theta_\mathrm{LS} \in \{110^\circ, 130^\circ, 140^\circ, 160^\circ\}$.}
   \label{fig: final spin surface}
\end{figure}

If we extend the aligned-spin condition for the transition point to precessing systems, then we might expect that the transition occurs at $\mathbf{S}^\parallel_\mathrm{f} = 0$. This assumption has been previously employed in models of precessing systems, for example, Ref.~\cite{Hamilton:2021pkf}. 
However, for a precessing system the orbital plane changes direction quite rapidly close to merger meaning it is ambiguous as to whether the sign of $\mathbf{S}^\parallel_\mathrm{f}$ truly indicates whether the final spin lies in the upper or lower hemisphere with respect to the orbital plane.
If we instead consider the orientation of the final spin relative to the optimal emission direction post-merger, rather than relative to the orbital plane prior to merger, then the orientation is described by the precession angle $\beta$. 
In this case, the final spin passes from the upper to lower hemisphere when $\beta = \pi/2$.
A second possible condition for the transition from prograde to retrograde frequencies occurs is therefore when $\cos\beta = 0$.

\begin{figure}[t]
   \centering
   \includegraphics[width=0.48\textwidth]{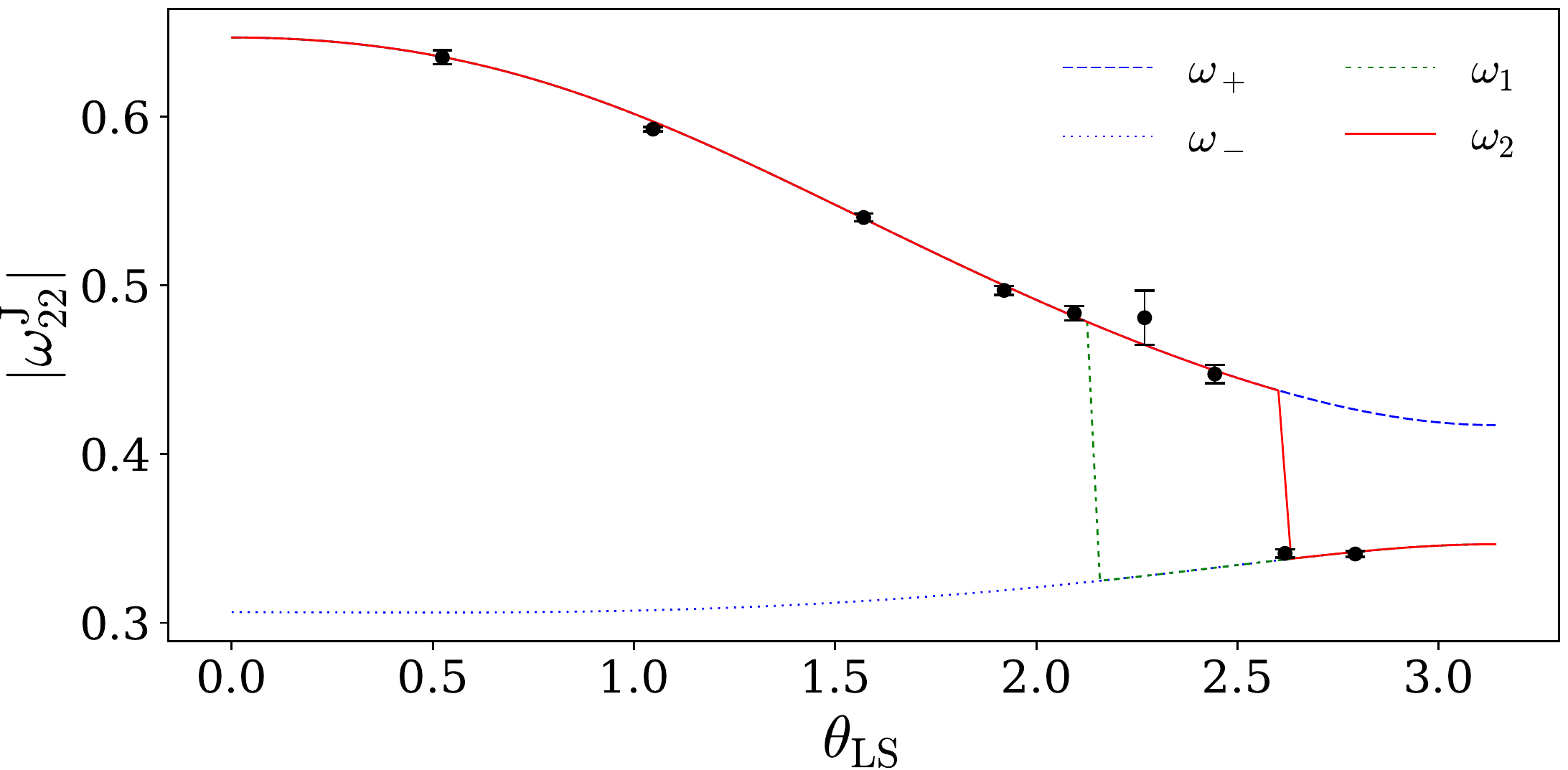} 
   \caption{Evolution of the mean ringdown frequency of the (2,2) multipole for binaries with $q=8$ and $\chi=0.8$ as the angle between the spin and orbital angular momenta varies. The black dots are the average ringdown frequency as calculated from numerical simulations. The error bars are obtained by varying the region over which the average is taken. The blue lines show the prograde (dashed) and retrograde (dotted) lines. The green dot-dashed line ($\omega_1$) shows how the mean ringdown frequency evolves across the parameter space as determined by the sign of $\chi_\mathrm{f}^\parallel$, as assumed in e.g. Refs.~\cite{bohe2016phenompv2,Khan:2018fmp}. The red solid line ($\omega_2$) shows the evolution of the mean ringdown frequency based on our assertion here that it is determined by the sign of $\cos\beta$. We can see that in the $\mathbf{J}$-frame there is a sharp transition between the prograde and retrograde modes.}
   \label{fig: J RD freq}
\end{figure}

Which of these approximations is most appropriate depends on the orientation of the perturbation of the final \bh{}. 
This perturbation is generated by the orbiting \bh{s} but its direction is described post-merger by the optimal emission direction.
In order to determine which of these approximations most accurately describes where the transition between prograde and retrograde frequencies occurs we studied six \nr{} simulations in a region of parameter space where both the prograde and retrograde frequencies are similarly excited.
These simulations are indicated in Fig.~\ref{fig: final spin surface}, which shows the magnitude of $\mathbf{S}^\parallel_\text{f}$, calculated using the fit presented in Ref.~\cite{Jimenez-Forteza:2016oae}, as a function of the mass ratio and aligned-spin component of the initial binary. 
For configurations that lie on the white line $\mathbf{S}^\parallel_\text{f} = 0$ and thus we might expect that, if the assumption that the orientation of the perturbation is consistent with that of the orbital plane just prior to merger, then the prograde and retrograde \qnm{s} are equally excited in these cases.
From this approximation we would expect the transition to occur for a value of $\theta_\mathrm{LS}$ between $120^\circ$ and $130^\circ$ for binaries with $q=8$ and $\chi=0.8$.
However, we can clearly see from Fig.~\ref{fig: J RD freq} that this is not the case and the sharp transition from prograde to retrograde frequencies does \emph{not} occur when $\mathbf{S}^\parallel_\text{f} = 0$.
The green dot-dashed line in this figure shows the predicted value of the mean ringdown frequency for binaries with $(q,\chi)=(8,0.8)$ and a range of spin inclination angles, assuming the final masses and spins as calculated using the fits in Ref.~\cite{Jimenez-Forteza:2016oae}. 
It jumps sharply from the prograde to the retrograde branch at the point where $\mathbf{S}_\mathrm{f}^\parallel=0$.
We can see from our \nr{} data (black dots) that the transition between the prograde and retrograde frequencies actually occurs for configurations much closer to the anti-aligned spin limit. 

\begin{figure}[t]
   \centering
   \includegraphics[width=0.48\textwidth]{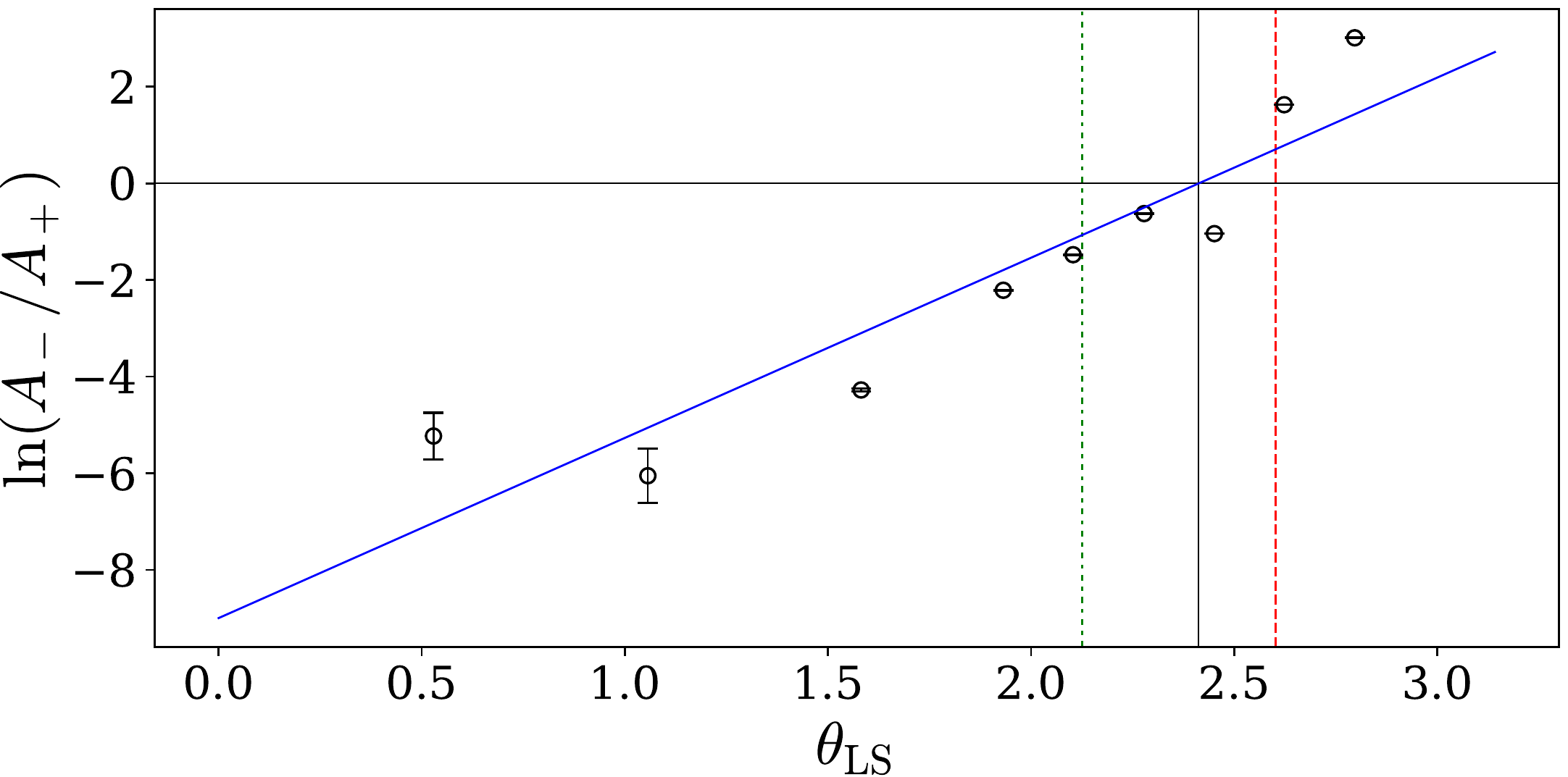} 
   \caption{Evolution of the ratio of the amplitudes and the prograde and retrograde (2,2) \qnm{s} for binaries with $q=8$ and $\chi=0.8$ as the angle between the spin and orbital angular momenta varies. The open circles are the ratio as calculated from time domain fits to numerical simulations~\cite{London:2014cma}. The time domain fitting was performed over three $30M$-long segments of data in the region from $30$ to $80M$ after merger. We plot the average value from these fits. The error bars are given by the maximum difference between the value over any given fitting region and the average value. The blue line gives a fit through this data, predicting that $A_+=A_-$ when $\theta_\mathrm{LS}=138^\circ$. For comparison, we show where the transition point from prograde to retrograde-dominated behaviour is predicted by the evolution of $\mathbf{S}^\parallel_\mathrm{f}$ (green dot-dashed) and of $\beta$ (red dashed line).}
   \label{fig: amplitude fit}
\end{figure}

This is consistent with what we see from looking at the amplitudes of the \qnm{s} themselves.
Predicting the exact point at which $A_+ = A_-$ is challenging since the number of \nr{} simulations available in this region of the parameter space is limited and the ratio of the prograde and retrograde amplitudes is not monotonic, as demonstrated in Fig.~\ref{fig: amplitude fit}.
Despite this, it is clear from Fig.~\ref{fig: amplitude fit} that, as seen in Fig.~\ref{fig: J RD freq}, this point occurs somewhere between $\theta_\mathrm{LS} = 140^\circ$ and $150^\circ$ for binaries with $q=8$, $\chi=8$.

\begin{figure}[t]
   \centering
   \includegraphics[width=0.48\textwidth]{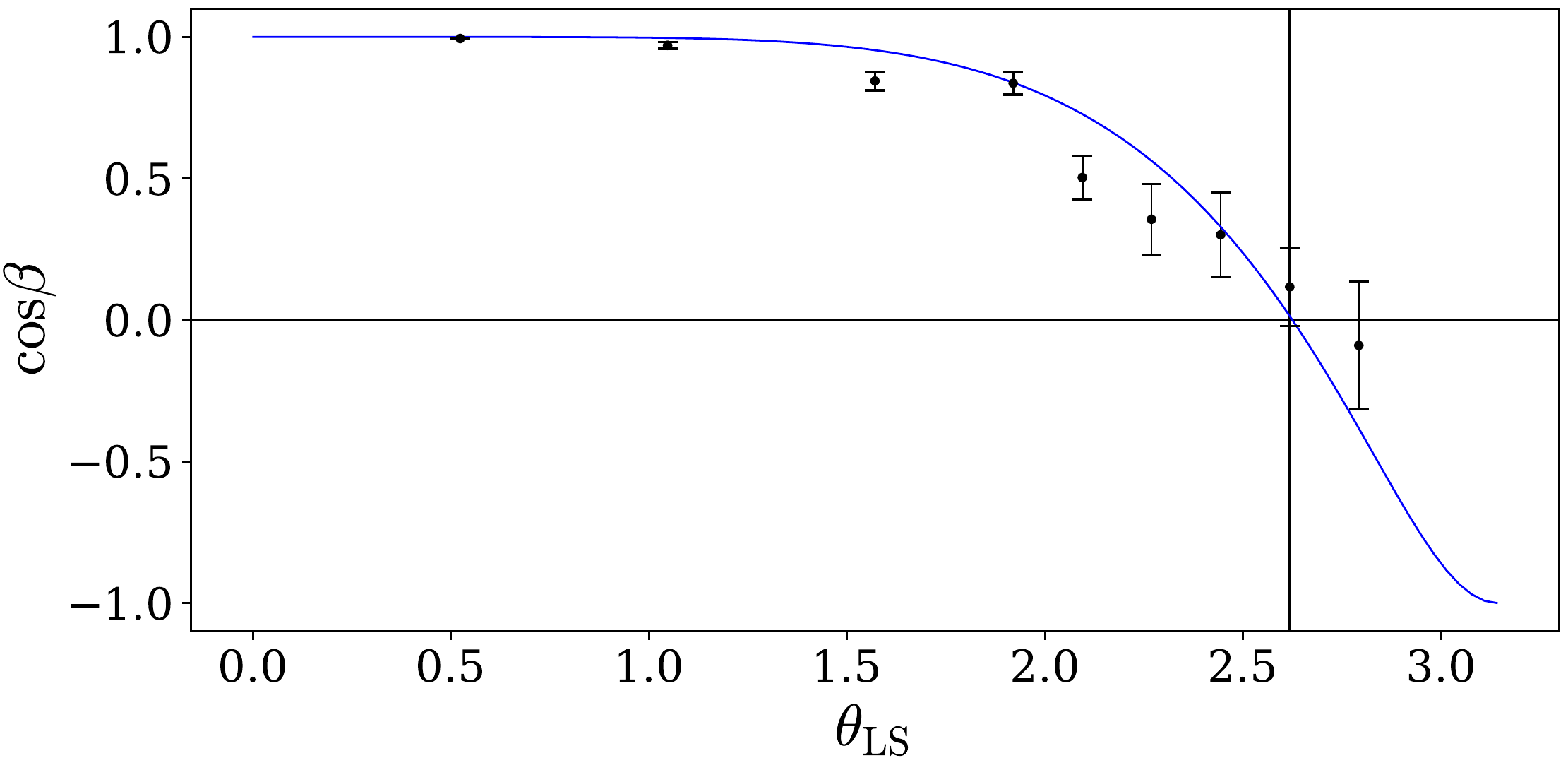} 
   \caption{Evolution of the ringdown value of $\beta$. The blue dots indicate the mean value of $\beta$ over $90M$ post-merger with an error bar indicating the amplitude of the oscillations in $\beta$. The blue line gives the the value of a fit for the frequency-domain ringdown value of $\beta$ from Ref.~\cite{Thompson:2023}. From the time domain values, it is difficult to estimate the exact point at which the $\cos\beta = 0$, however it is clear that it occurs in the region around $150^\circ$--$160^\circ$. The frequency domain fit predicts that this will occur when $\theta_\mathrm{LS}=150^\circ$.}
   \label{fig: final beta evolution}
\end{figure}

If we now consider the orientation of the final spin with respect to the direction of the perturbation through ringdown (the optimal emission direction) then we find it is much more consistent with our \nr{} data.
Fig.~\ref{fig: final beta evolution} shows the evolution of the ringdown value of $\beta$ across the parameter space, and we can see that although it is not clear exactly where $\cos\beta$ passes through zero it certainly does not happen around $120^\circ$ and occurs closer to $150^\circ$, which is more consistent with what we see in Figs.~\ref{fig: J RD freq} and~\ref{fig: amplitude fit}. This approximation is therefore much more accurate in determining the point at which the transition from prograde to retrograde frequencies occurs. 

Having understood the behaviour of the mean value of the ringdown frequency, we can also examine the validity of our prediction of the oscillations present in the angular frequency, as described by Eq.~(\ref{eqn: angular freq J}).
In order for the oscillations in the angular frequency to be visible in our \nr{} data both the prograde and retrograde \qnm{s} must be reasonably excited (so the amplitude of the oscillations are visible above the level of numerical noise in the data). This limits us to a very small region in the currently explored precessing-binary parameter space where this phenomenon can be studied. 
It is configurations in this region of parameter space for which we most clearly see the oscillations in the angular frequency of the individual multipoles. 

In Fig.~\ref{fig: oscillatory angular frequency} we compare the angular frequency of the (2,2) multipole in the $\mathbf{J}$-frame as calculated from the \nr{} data with the prediction from Eq.~(\ref{eqn: angular freq J}). We consider two cases; in the top panel we consider the case where the prograde frequency dominates while in the second panel we consider the case where the retrograde frequency dominates.  The good agreement between the numerical data and our prediction strongly suggests that it is the superposition of these two \qnm{s} that accounts for these oscillations in the numerical data rather than numerical noise. We can see that the mean value is determined by the dominant multipole as predicted by Eq.~(\ref{eqn: pro retro qnm selection}), the frequency of the oscillations is determined by the difference between the two \qnm{} frequencies and the amplitude of the oscillations is determined by the ratio of the amplitudes of the prograde and retrograde modes.

In this figure, we consider the angular frequency as predicted by Eq.~(\ref{eqn: angular freq J}) using values for the \qnm{} amplitudes obtained from fitting a series of damped sinusoids to the individual spherical multipoles, as described in Sec.~\ref{sec: NR}. There is some degree of uncertainty in this procedure. 
We can alternatively find values for the ratio of the amplitudes by fitting Eq.~(\ref{eqn: angular freq J}) to the numerical data.
This gives results very close to those obtained by the independent amplitude fitting. 
This fit is much simpler than the damped sinusoid fit as it has fewer free parameters. 
However, it does not provide us with the absolute values of the amplitudes.

\begin{figure}[t]
   \centering
   \includegraphics[width=0.48\textwidth]{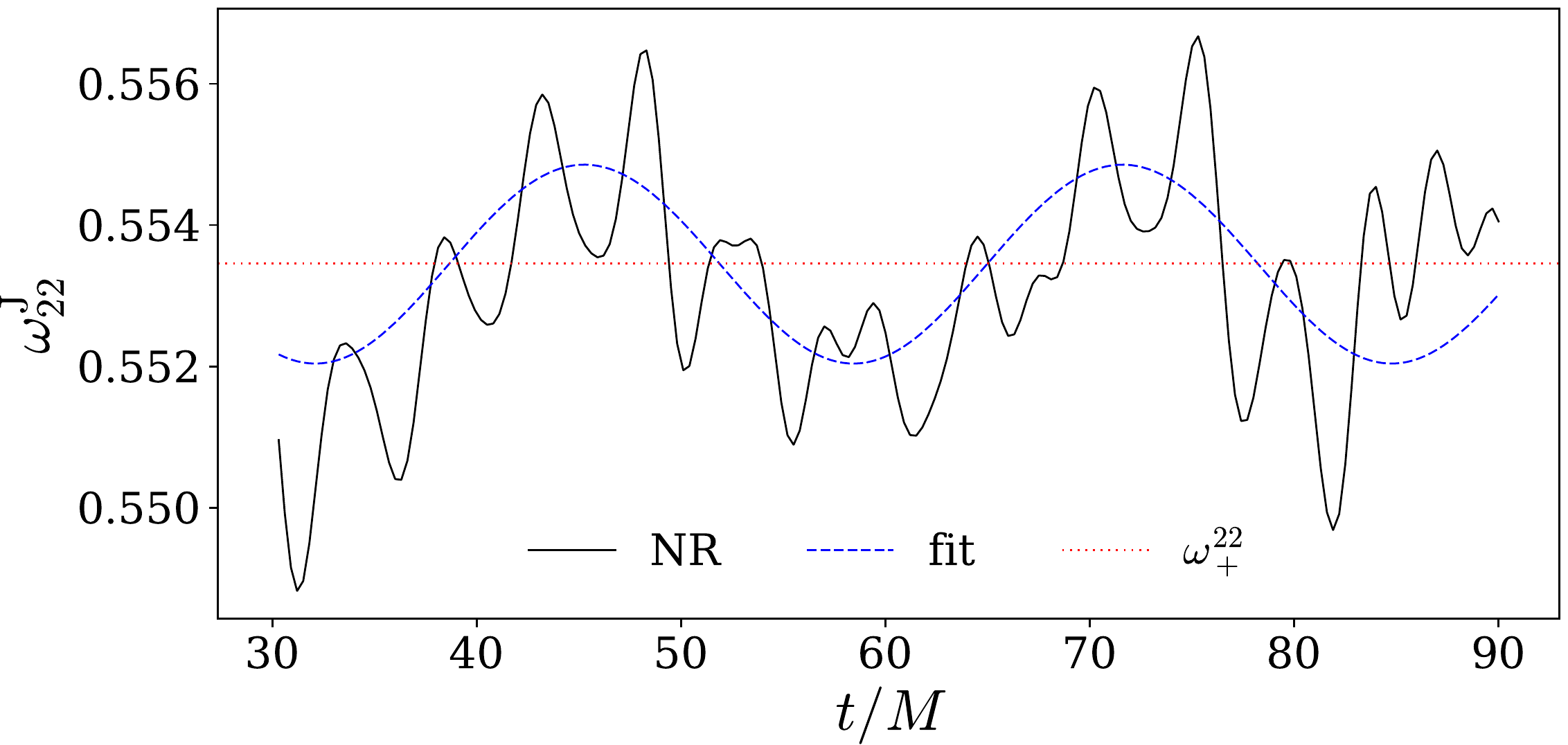}
   \caption{The time domain angular frequency of the (2,2) multipole in the $\mathbf{J}$-frame for a non-spinning equal mass binary (black). The oscillations present in the angular frequency of this non-spinning case are two orders of magnitude smaller in amplitude than in the precessing cases shown in Fig.~\ref{fig: oscillatory angular frequency}. The lower frequency oscillations are consistent with mode mixing with the (3,2) multipole with a ratio $A_{22} : A_{32}$ of $169:1$ (blue dashed line). The higher frequency oscillations are artefacts introduced by noise or gauge artefacts.
   }
   \label{fig: angular frequency nonspinning}
\end{figure}

In Fig.~\ref{fig: angular frequency nonspinning} we show the angular frequency of the (2,2) multipole for an equal mass non-spinning binary.
As we expect, in this case only the prograde frequencies are excited and so we see no oscillations due to the superposition of the prograde and retrograde \qnm{s}. 
We do see weak low frequency oscillations due to mixing of the (2,2) and (3,2) \qnm{s} (see discussion in the following section) as well as higher frequency oscillations due to noise artefacts and gauge effects.
These higher frequency oscillations are inconsistent with a superposition of \qnm{s} and so cannot be accounted for in this fashion.
The oscillations visible in this case are $\mathcal{O}(10^2)$ times weaker than those seen due to the superposition of the prograde and retrograde frequencies in Fig.~\ref{fig: oscillatory angular frequency}, showing that neither the contribution of the (3,2) \qnm{} nor the gauge effects have a strong effect on the signal.

In what follows we will consider the multipoles in the inertial $\mathbf{J}$-frame to have a single frequency only, given by the mean value, and neglect the effect of the oscillations.

\subsection{Spherical harmonic picture}\label{sec: spherical mode mixing}

When considering the \gw{s} from a perturbed Kerr \bh{} in a spherical harmonic basis rather than the spheroidal basis we get mixing in some of the subdominant multipoles. This mixing occurs between spheroidal multipoles with the same $m$ value but different $\ell$ values. Of the five most important multipoles; the (2,2), (2,1), (3,3), (3,2) and (4,4) multipoles, mixing is apparent in only the (3,2) multipole. While in theory this multipole couples to all other $m=2$ multipoles, we need only consider the contribution from the (2,2) multipole. The multipoles in the two bases are then related via the linear transformation~\cite{Berti:2014fga,Garcia-Quiros:2020qpx}
\begin{align}
   \begin{pmatrix} 
      \psi_{4,22} \\
      \psi_{4,32} \\
   \end{pmatrix}
   = {}& 
   \begin{pmatrix} 
      \alpha_{222} & \alpha_{232} \\
      \alpha_{322} & \alpha_{332} \\
   \end{pmatrix}
   \begin{pmatrix} 
      \psi^S_{4,22} \\
      \psi^S_{4,32} \\
   \end{pmatrix},
\end{align}
where $S$ indicates the multipoles in the spheroidal harmonic basis and $\alpha_{\ell \ell' m}$ are the mixing co-efficients. This gives an expression for the spherical harmonic multipoles in the same form as Eq.~(\ref{eqn: pro retro sum}). Following the same calculation as described above, we expect the mean angular frequency of the (3,2) multipole in the spherical harmonic basis to be given by 
\begin{equation}\label{eqn: mm qnm selection}
    \langle \dot\phi_{32}\rangle =
    \begin{cases}
      \omega_{22}, & \alpha_{322}A^S_{22} > \alpha_{332}A^S_{32} \\
      \omega_{32}, & \alpha_{322}A^S_{22} < \alpha_{332}A^S_{32} \\
      \frac{1}{2}\left(\omega_{22}+\omega_{32}\right), & \alpha_{322}A^S_{22} = \alpha_{332}A^S_{32}
    \end{cases},
  \end{equation}
where $A^S_{\ell' m}$ are the amplitudes of the multipoles in the spheroidal harmonic basis.

\begin{figure}[t]
   \centering
   \includegraphics[width=0.48\textwidth]{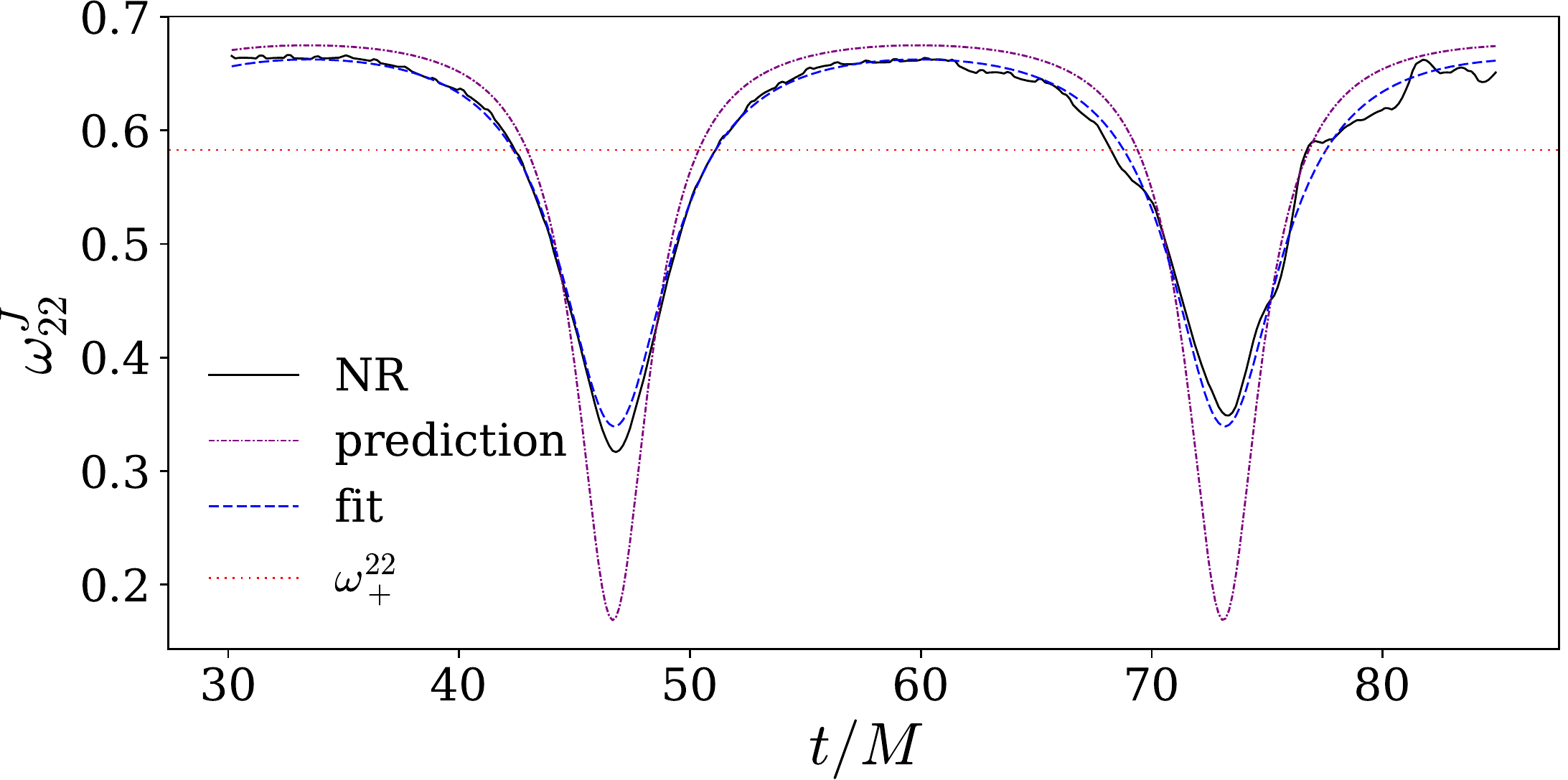} \\
   \includegraphics[width=0.48\textwidth]{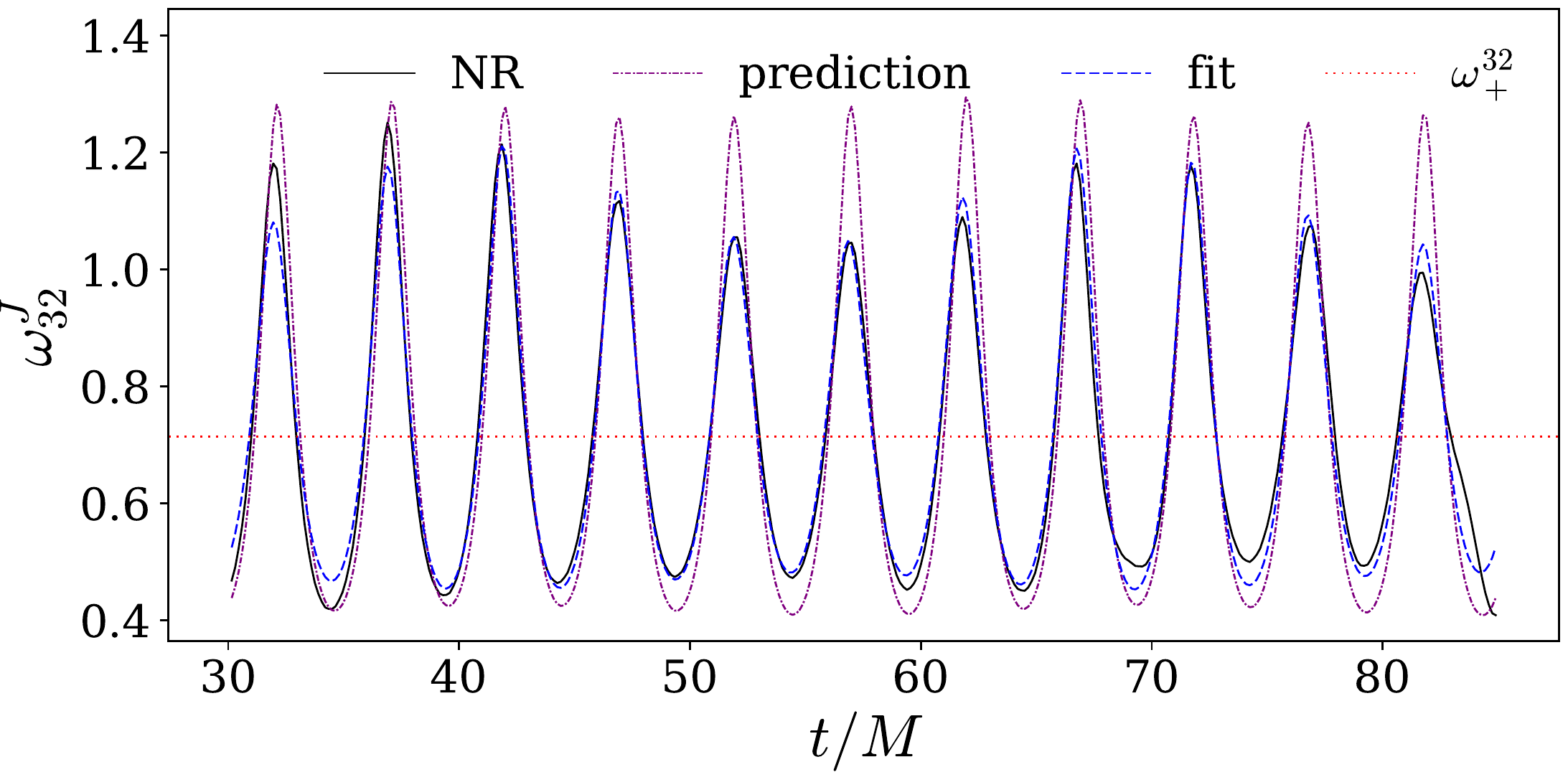} 
   \caption{Effect of mixing on the $\mathbf{J}$-frame angular frequency of the (3,2) multipole. The top panel shows the case CF21-6, while the bottom panel shows CF-79. 
   In the top panel we see oscillations due to the superposition of the (3,2) and (2,2) \qnm{s} only.
   From fitting damped sinusoids to the spherical multipoles (purple dot-dashed line) we find the ratio $A_{22}^+:A_{32}^+ = 1:0.636$. From fitting the angular frequency (blue dashed line) directly we find $A_{22}^+:A_{32}^+ = 1:0.506$.
   In the lower panel we see the oscillations provided by the superposition of the three three dominant \qnm{s}: the envelope comes from the mixing between the (3,2) and the (2,2) multipoles while the faster oscillations come from the superposition of the prograde and retrograde modes. 
   Fitting the multipoles themselves gives the relative amplitude of the \qnm{s} as $A_{22}^+:A_{22}^-:A_{32}^+:A_{32}^- = 1:0.440:23.7:7.30$  while fitting the angular frequency gives $1:0.401:18.3:4.44$.
   In both cases it is clear that the mean frequency is given by the dominant \qnm{} frequency, indicated by the red dotted line.
   }
   \label{fig: mode mixing}
\end{figure}

As with the superposition of the prograde and retrograde frequencies, the superposition of the different spheroidal harmonic multipoles causes oscillations in the spherical harmonic multipole. The oscillations induced by this mixing can be seen in Fig (\ref{fig: mode mixing}). 
In the top panel we see a case where the amplitude of any prograde modes greatly exceeds that of any retrograde modes, so any oscillations that might be due to their superposition are not apparent. 
Consequently, the oscillations due to the superposition of the (2,2) and (3,2) \qnm{s} are clearly visible. 
In the lower panel, we consider a case where the amplitudes of the prograde and retrograde modes are comparable and we do indeed see oscillations due to superposition of these modes.
Here, the oscillations due to the superposition of the (2,2) and (3,2) \qnm{s} can be seen in the envelope around the oscillations.
Similar to when looking at the superposition of the prograde and retrograde modes in the (2,2) multipole, we can get the \qnm{} amplitudes by either fitting a series of damped sinusoids to the waveform or by fitting the angular frequency data itself.
As can be seen, the two methods give good agreement in the relative amplitudes but the small differences have a clear effect on the magnitude of the oscillations in the angular frequency.
In both cases shown in Fig.~\ref{fig: mode mixing}, we can clearly see that the mean behaviour is given by the frequency of the dominant \qnm{}, even in the case where we have a superposition of not only the (3,2) and (2,2) multipoles but also of the prograde and retrograde modes. 
As discussed in the previous section, it is this mean value which is of greatest interest.
In the rest of the paper we will therefore ignore the effect of the oscillations and instead consider the multipole to have a single frequency, given by the mean value.

For the results presented in the following sections, we will consider the spherical harmonic decomposition. 
For all multipoles considered in this paper besides the (3,2) multipole, the mean value of the ringdown frequency is unaffected by the mode mixing.
For the (3,2) multipole, mode mixing means that at times the (3,2) \qnm{} dominates and at others the (2,2) \qnm{} dominates.
This determines whether the mean frequency is equal to the (2,2) or the (3,2) \qnm{} frequency.

\section{Co-precessing frame behaviour}\label{sec: cp frame}

\begin{figure*}[htbp]
   \centering
   \includegraphics[width=\textwidth]{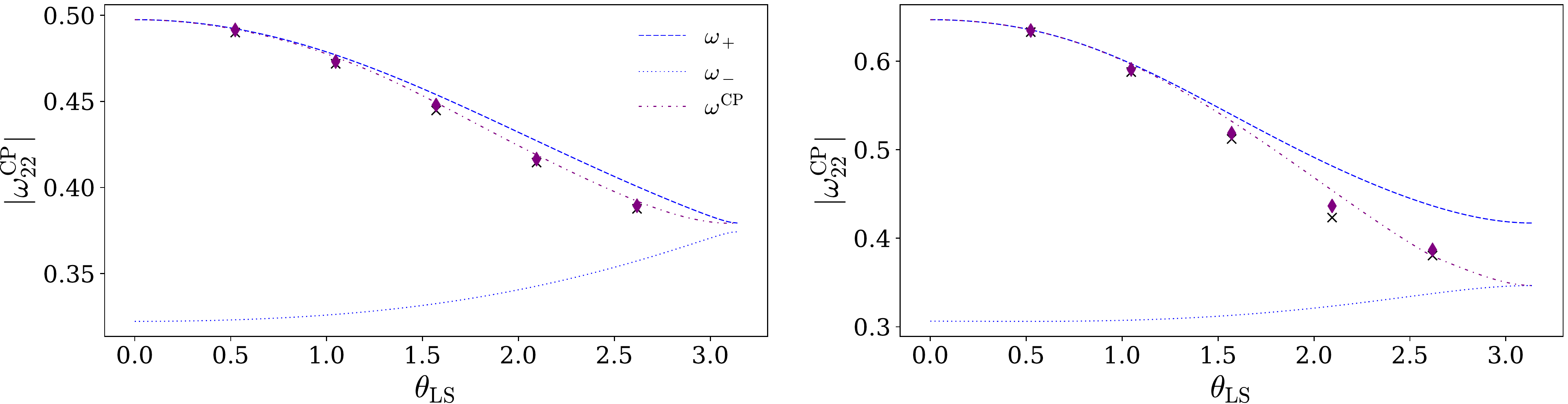} \\
   \includegraphics[width=\textwidth]{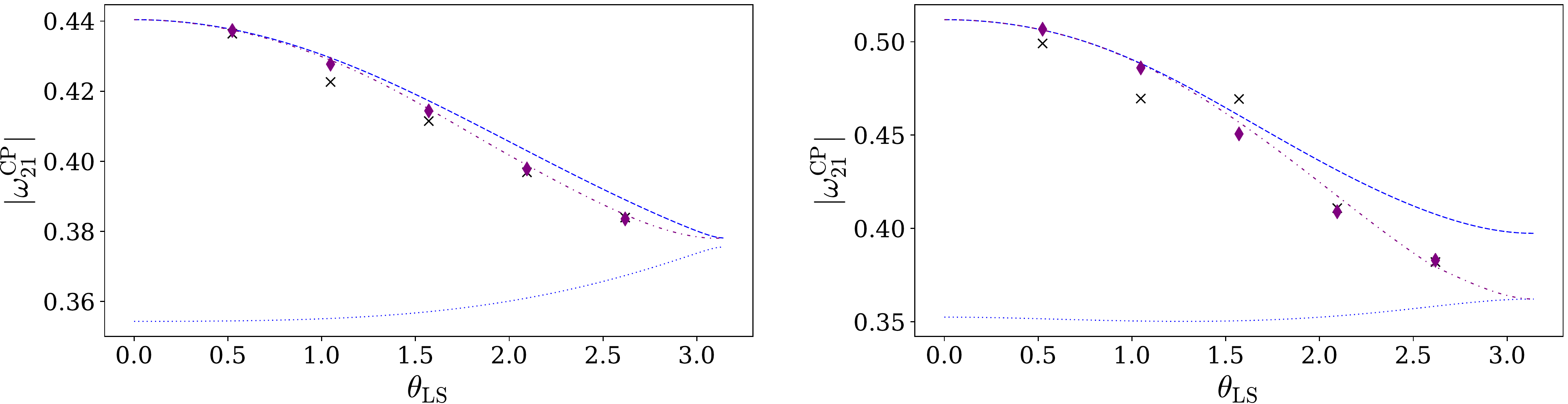} \\
   \includegraphics[width=\textwidth]{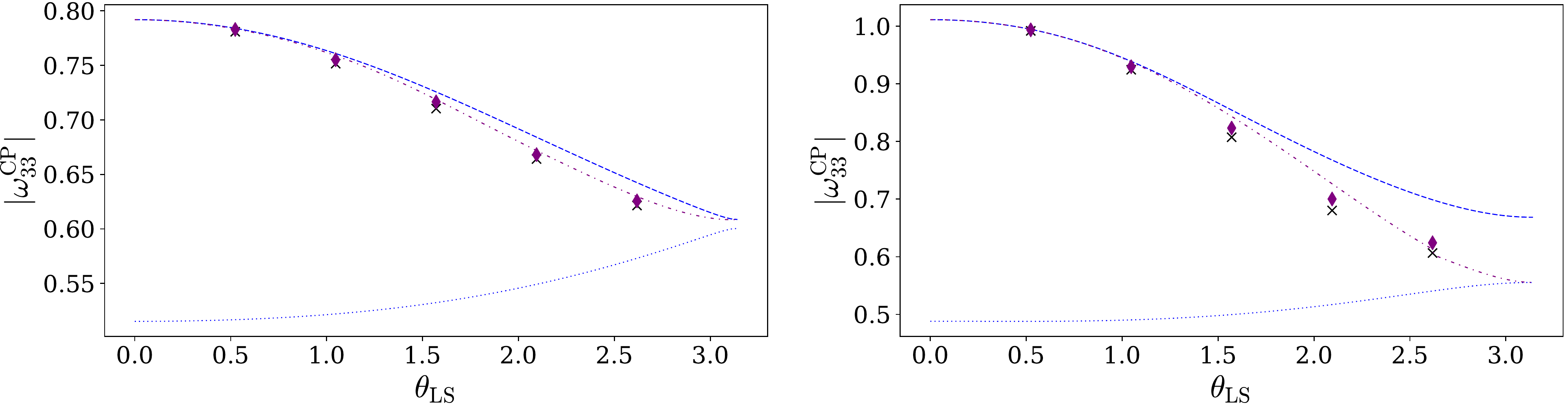} \\
   \includegraphics[width=\textwidth]{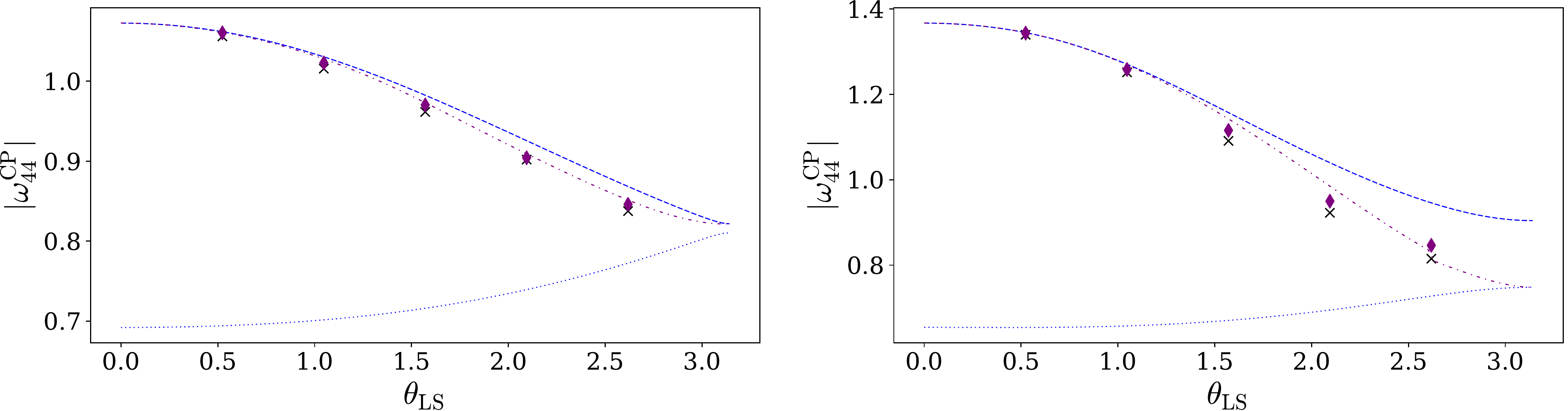} 
   \caption{A comparison of the prediction of the effective ringdown frequency given by Eq.~(\ref{eqn: HOM expression}) (purple diamonds) with the mean value of the frequency of the multipole in the co-precessing frame as calculated from the \nr{} data (black crosses). The blue dashed and dotted lines show the absolute value of the prograde and retrograde frequencies for the relevant multipole. The purple dot-dashed line shows the prediction of the effective ringdown frequency using a value for $\beta$ based on the fits in Ref.~\cite{Thompson:2023}. The panels are arranged in the order (2,2), (2,1), (3,3) and (4,4). All systems considered here are $q=8$. The left hand column shows $\chi=0.6$ while the right hand column shows $\chi=0.8$.}
   \label{fig: RD beta comparison}
\end{figure*}

We now consider the ringdown signal in the co-precessing frame, which has been used in the construction of many precessing-binary signal
models~\cite{Hannam:2013oca, Khan:2018fmp, Khan:2019kot, Pratten:2020ceb, Pan:2013rra, Taracchini:2013rva, Ossokine:2020kjp, Estelles:2021gvs,Blackman:2017pcm,Varma:2019csw}.
The results observed in this section have already been employed in the model presented in Ref.~\cite{Hamilton:2021pkf}.
We work in the spherical harmonic basis. The multipoles in this frame are a superposition of all the multipoles in the $\mathbf{J}$-frame with the same $\ell$ value. If the $\mathbf{J}$-frame multipoles with the same $\ell$ are of comparable magnitude then we will see noticeable oscillations in the co-precessing frame multipoles. Otherwise, we just see the mean behaviour.

\subsection{Mean behaviour}

The co-precessing multipoles are given in terms of the $\mathbf{J}$-frame multipoles by Eq.~(\ref{eqn: multipole rotation}). Writing $\psi_{4,\ell m} = A_{\ell m}e^{-i\phi_{\ell m}} = A_{\ell m}\left( \cos\phi_{\ell m} - i \sin\phi_{\ell m}\right)$ we get the following expression for the $(\ell, m')$ multipole in the co-precessing frame:
\begin{align} 
   \psi'_{4,\ell m'} = {}& e^{-im'\gamma} 
   \sum_{m=-\ell}^\ell A_{\ell m}d^\ell_{m' m}\left(-\beta\right)  \nonumber \\
   {}& \quad \times \left[
   \cos\left(\phi_{\ell m}-m\alpha\right) - i
   \sin\left(\phi_{\ell m}-m\alpha\right)\right].
\end{align}
From this we can see that the phase of the $(\ell,m')$ multipole in the co-precessing frame is given by
\begin{align} \label{eqn: full cp phase}
   \tan\left(\phi'_{\ell m'} + m' \gamma\right) = {}&
   \frac{\sum_{m=-\ell}^\ell A_{\ell m} d^\ell_{m' m} \sin\left(\phi_{\ell m}-m\alpha\right)}
   {\sum_{m=\ell}^\ell A_{\ell m} d^\ell_{m' m} \cos\left(\phi_{\ell m}-m\alpha\right)}.
\end{align}

The effect of the rotation by $\alpha$ is to ensure that the magnitude of the phase of each of these terms is the same (i.e. $\phi_{\ell m} - m\alpha = \phi_{\ell m'} - m'\alpha$). 
We note that the sign of the phase of the negative $m$ multipoles is opposite to that of the positive $m$ multipoles. Eq.~(\ref{eqn: full cp phase}) then becomes
\begin{align} \label{eqn: full cp phase simplified}
   {}&\tan\left(\phi'_{\ell m'} + m' \gamma\right) = \nonumber \\
   {}& \quad \frac{\sum_{m=0}^\ell A_{\ell m} d^\ell_{m' m} - \sum_{m=-\ell}^{-1} A_{\ell m} d^\ell_{m' m}}
   {\sum_{m=-\ell}^\ell A_{\ell m}d^\ell_{m' m}}
   \tan\left(\phi_{\ell m'}-m'\alpha\right).
\end{align}

Making use of the assumptions that (i) the value of $\beta$ through ringdown is constant and (ii) $\alpha$ and $\gamma$ are linear through ringdown, we can rewrite Eq.~(\ref{eqn: full cp phase simplified}) as 
\begin{align} 
   \tan\left(\phi'_{\ell m'} + m' \gamma\right) = {}& b\tan\left(at\right),
\end{align}
where
\begin{align} 
   a = {}& \dot\phi_{\ell m'}-m'\dot\alpha = \omega_{\ell m'}-m'\dot\alpha, \\
   b = {}& \frac{\sum_{m=0}^\ell A_{\ell m} d^\ell_{m' m} - \sum_{m=-\ell}^{-1} A_{\ell m} d^\ell_{m' m}}
   {\sum_{m=-\ell}^\ell A_{\ell m}d^\ell_{m' m}}, \label{eqn: b coeff}
\end{align}
are constants. 
Differentiating with respect to time we find
\begin{align}\label{eqn: full cp freq}
   \dot\phi'_{\ell m'} + m'\dot\gamma  = {}& 
   \frac{ab\left(1+\tan^2\left(at\right)\right)}{1+b^2 \tan^2\left(at\right)}.
\end{align}
The average value of the frequency of the $(\ell, m')$ multipole in the co-precessing frame $\omega'_{\ell m'} = \langle \dot\phi'_{\ell m'} \rangle$ is therefore given by
\begin{align}
   \omega'_{\ell m'} = {}& \omega_{\ell m'} - m'\left(\dot\alpha + \dot\gamma\right) \nonumber \\
    = {}& \omega_{\ell m'} - m'(1 - |\cos\beta|)\dot\alpha, \label{eqn: cp expression alpha}
\end{align}
where in the second line we have used the minimal rotation condition given by Eq.~(\ref{eqn: mrc}).
We take the absolute value of $\cos\beta$ since we define $\beta$ to be the angle between the total angular momentum and the optimal emission direction, which has no preferred direction. 
We therefore require the minimal rotation condition to be symmetric about $\beta=\pi/2$.

Employing Eq.~(\ref{eqn: alphadot}) we get the following final expression for the effective ringdown frequency of the $(\ell,m)$-multipole,
\begin{align} \label{eqn: HOM expression}
   \omega'_{\ell m'} = {}& \omega_{\ell m'} - m'(1-|\cos\beta|)\left(\omega_{22} - \omega_{21}\right).
\end{align}
This depends only on the QNM frequencies and the ringdown value of $\beta$.

This expression gives us a simple way to obtain the correct ringdown frequency in the co-precessing frame from the results from perturbation theory.
This is useful for modelling precessing systems since we can obtain the ringdown frequencies for each of the multipoles in the co-precessing frame based purely on knowledge of the initial binary and the remnant black hole, without additional modelling. 
Indeed, this expression has already been employed for the $\ell \neq m$ multipoles in Ref.~\cite{Thompson:2023}.

Since this derivation has been performed entirely in the context of a spherical harmonic decomposition, it should be noted that $\omega_{\ell m'}$ must be the correct ringdown frequency of the spherical harmonic multipole in the $\mathbf{J}$-frame. That is, not only must it be the correct choice of the prograde or retrograde frequency as determined by Eq.~(\ref{eqn: pro retro qnm selection}), in the case of mixing it must also be the correct \qnm{} frequency as determined by Eq.~(\ref{eqn: mm qnm selection}).

One great advantage of this formula is that, given a model for $\beta$, one can predict the effective ringdown frequency of each of the higher order multipoles in the co-precessing frame. This means one needs only to produce a model for $\beta$ rather than for each individual multipole of interest. 

\begin{figure*}[!t]
   \centering
   \includegraphics[width=\textwidth]{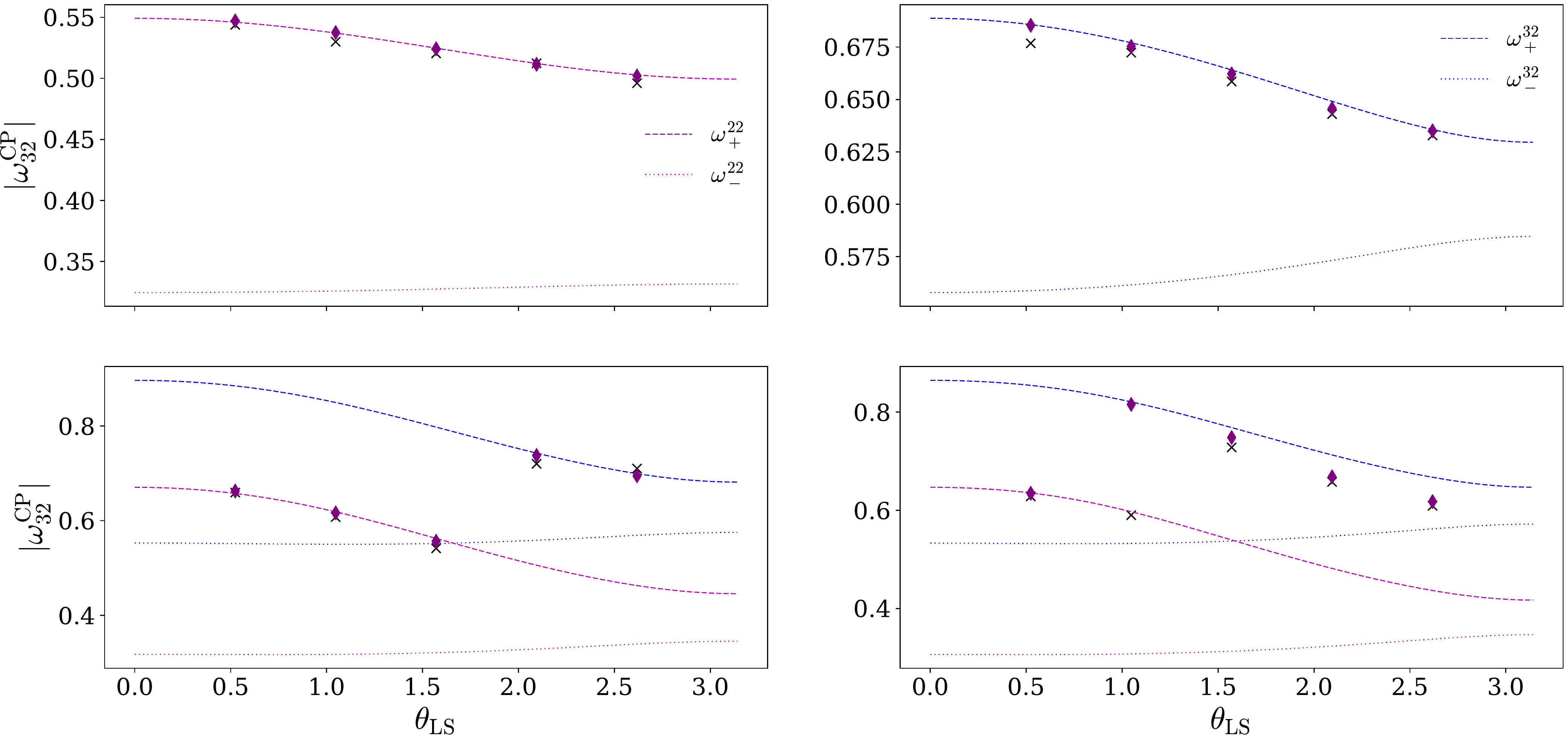}
   \caption{A comparison of the prediction of the effective ringdown frequency given by Eq.~(\ref{eqn: HOM expression}) for the (3,2) multipole (purple diamonds) with the mean value of the frequency in the co-precessing frame as calculated from the \nr{} data (black crosses). The dashed lines show the prograde and retrograde frequencies for the (2,2) multipole (magenta) and the (3,2) multipole (blue). The top row shows systems with $\chi=0.2$ while in the bottom row we consider systems with $\chi=0.8$. The left hand column shows $q=2$ configurations and the right hand column shows $q=8$.}
   \label{fig: 32 RD beta comparison}
\end{figure*}

The validity of the expression for the effective ringdown frequency given in Eq.~(\ref{eqn: HOM expression}) is demonstrated in Figs.~\ref{fig: RD beta comparison} and \ref{fig: 32 RD beta comparison}. 
The mean value of the angular frequency in the co-precessing frame (averaged over the range $40$--$90M$ after merger) is represented by the black crosses.
For comparison, the result of Eq.~(\ref{eqn: HOM expression}) for each of our \nr{} cases is represented by the purple diamonds.
We calculated the \qnm{} frequency for use in Eq.~(\ref{eqn: HOM expression}) using Leaver's method for the final mass and spin of the black hole given in the \nr{} metadata. We used the mean value of ringdown $\beta$ taken from the numerical data between $40$ and $90M$ after merger. We can see that the result of Eq.~(\ref{eqn: HOM expression}) agrees very well with the value of the angular frequency from the \nr{} data, thus demonstrating that this is a good approximation of the effective ringdown frequency. 
For further comparison, we show the result of Eq.~(\ref{eqn: HOM expression}) across the parameter space in the purple dot-dashed line. To obtain this prediction, we used the final mass and spin calculated using the fits for aligned binaries in Ref.~\cite{Jimenez-Forteza:2016oae} as well as Eq.~(\ref{eqn: final spin}) to account for the precessing spin and the final value of $\beta$ taken from frequency domain fits presented in Ref.~\cite{Thompson:2023}.

Crucially, in Fig.~\ref{fig: RD beta comparison} we can see that the discontinuity in the ringdown frequency as we travel across the parameter space has disappeared in the co-precessing frame. 
The effect of transforming into the co-precessing frame is clearest for cases with large mass ratio, spin magnitude and spin inclination angle, so once again we are focussed in this corner of the parameter space. 
We examine how the effective ringdown frequency changes as we travel from aligned to anti-aligned-spin systems. 
As mentioned previously, for aligned-spin systems the final spin is always positive (i.e. aligned with the axis of the perturbation).
For anti-aligned systems, the final spin can be either positive or negative depending on the configuration of the initial binary.

For anti-aligned cases where the final spin is positive, we will simply have the prograde ringdown frequency. 
Here there is a smooth deviation from and return to the prograde value when travelling across the parameter space, as seen for the $q=8$, $\chi=0.6$ systems shown in the left hand column of Fig.~\ref{fig: RD beta comparison}.
For anti-aligned cases where the final spin is negative, the effective ringdown frequency transitions smoothly between the prograde frequency for aligned-spin systems and the retrograde frequency for anti-aligned systems. 
This is the case for the $q=8$, $\chi=0.8$ systems shown in the right hand column of the figure.

This smooth transition occurs provided we have correctly understood when the transition between prograde and retrograde frequencies occurs in the $\mathbf{J}$-frame. 
At this point the shifted value of the prograde and retrograde frequencies in the co-precessing frame are equal, i.e.
\begin{align}
   \omega'^+_{\ell m} 
   = {}& -\omega'^-_{\ell m},
\end{align}
where $\omega'^\pm_{\ell m}$ are each given by Eq.~(\ref{eqn: HOM expression}).
We see from Fig.~\ref{fig: RD beta comparison} that, for the $\ell=2$ multipoles, this occurs when $\cos\beta=0$.
This agrees with the conclusion drawn in Sec.~\ref{sec: J frame spheroidal} that $\cos\beta$ is a good indicator of the dominant \qnm{} frequency.
For the higher order multipoles the transition is not quite so smooth. This may be due, at least in part, to the approximation for $\dot{\alpha}$ made in Eq.~(\ref{eqn: alpha approx}) breaking down with the inclusion of higher order multipoles. 

In the case of the (3,2) spherical multipole shown in Fig.~\ref{fig: 32 RD beta comparison}, we can see that the mean frequency is given by that of either the (3,2) \qnm{} frequency or the (2,2) \qnm{} frequency depending on which one dominates in the co-precessing frame, as discussed in Sec.~\ref{sec: spherical mode mixing}. The transition between these two frequencies is discontinuous at the point where the amplitude of the contributions from the two \qnm{s} to the spherical multipole are equal. If one were to consider the multipoles decomposed in the spheroidal harmonic basis, as is optimum when considering the (3,2) multipole (see. e.g. Ref.~\cite{Garcia-Quiros:2020qpx}), then the discontinuity would disappear and we would see the same trend as for the other multipoles shown in Fig.~\ref{fig: RD beta comparison}. 

We can see that for systems with low mass ratio (left hand column) the (2,2) \qnm{} tends to dominate while for systems with higher mass ratio the (3,2) \qnm{} dominates. This is as expected since for higher mass ratio systems, the higher order multipoles are excited more strongly. As the angle between the orbital and spin angular momenta of the binary increases, the (3,2) \qnm{} also starts to dominate (and vice versa). This is because more anti-aligned systems will result in a lower final spin magnitude and thus a higher contribution from the (3,2) \qnm{} relative to the contribution of the (2,2) since $\alpha_{332}$ increases relative to $\alpha_{322}$. This trend is most marked for systems where the inspiraling binary has a higher spin magnitude (the bottom row).

The dominant multipole used to calculate the effective ringdown frequency using Eq.~(\ref{eqn: HOM expression}) (purple diamonds in Fig.~\ref{fig: 32 RD beta comparison}) was determined using time domain fitting of the ringdown (3,2) spherical multipole in the co-precessing frame.
This is compared with the mean value of the angular frequency of the (3,2) multipole in the co-precessing frame (black crosses).
As can be seen from Fig.~\ref{fig: 32 RD beta comparison}, the prediction agrees well with the data for all cases except that with $q=8$, $\chi=0.8$ and $\theta_\text{LS}=60^\circ$.
In this case, the time domain fitting shows that the contribution of the (2,2) and (3,2) \qnm{s} to the (3,2) multipole are nearly equal. The time domain fitting favours the (3,2) dominating but the angular frequency implies the (2,2) may dominate.
However, the angular frequency data itself is fairly noisy.
It is therefore ambiguous as to which of these results is correct.

\subsection{Oscillatory behaviour}

As well as considering the mean behaviour of the spherical harmonic multipoles in the co-precessing frame we can also study the oscillations in the frequency, as described by Eq.~(\ref{eqn: full cp freq}).

To get a sense of exactly how these oscillations arise as a consequence of the rotations between frames, we consider only the $\ell=2$ multipoles, in the co-precessing frame defined considering the $\ell=2$ multipoles only.

For this example we assume that the (2,2) multipole is sufficiently dominant in the co-precessing frame that we can treat the other multipoles as negligible. This assumption is obviously very approximate for higher mass ratio systems. The ratio of the amplitudes of the multipoles in the $\mathbf{J}$-frame can therefore be found from Eq.~(\ref{eqn: dominant multipole transformation}). We therefore re-write Eq.~(\ref{eqn: b coeff}) as 
\begin{align}
   b = {}& \frac{2\cos\beta + 4\tan\frac{\beta}{2} \, \sin\beta 
   + \frac{3}{2}\tan^2\frac{\beta}{2} \, \sin^2\beta}
   {1+\cos^2\beta + 4\tan\frac{\beta}{2} \, \sin\beta\cos\beta
   + \frac{3}{2}\tan^2\frac{\beta}{2} \, \sin^2\beta}.
\end{align}
We can re-write Eq.~(\ref{eqn: HOM expression}) to find the expression for the mean frequency of the co-precessing (2,2) multipoles is given by
\begin{align}\label{eqn: l2 mean beta}
    \omega' = {}& \left(\left(2\kappa-1\right) + 2\left(1-\kappa\right)\cos\beta\right)\omega_{22},
\end{align}
where $\kappa = \frac{\omega_{21}}{\omega_{22}}$.

\begin{figure}[t]
   \centering
   \includegraphics[width=0.48\textwidth]{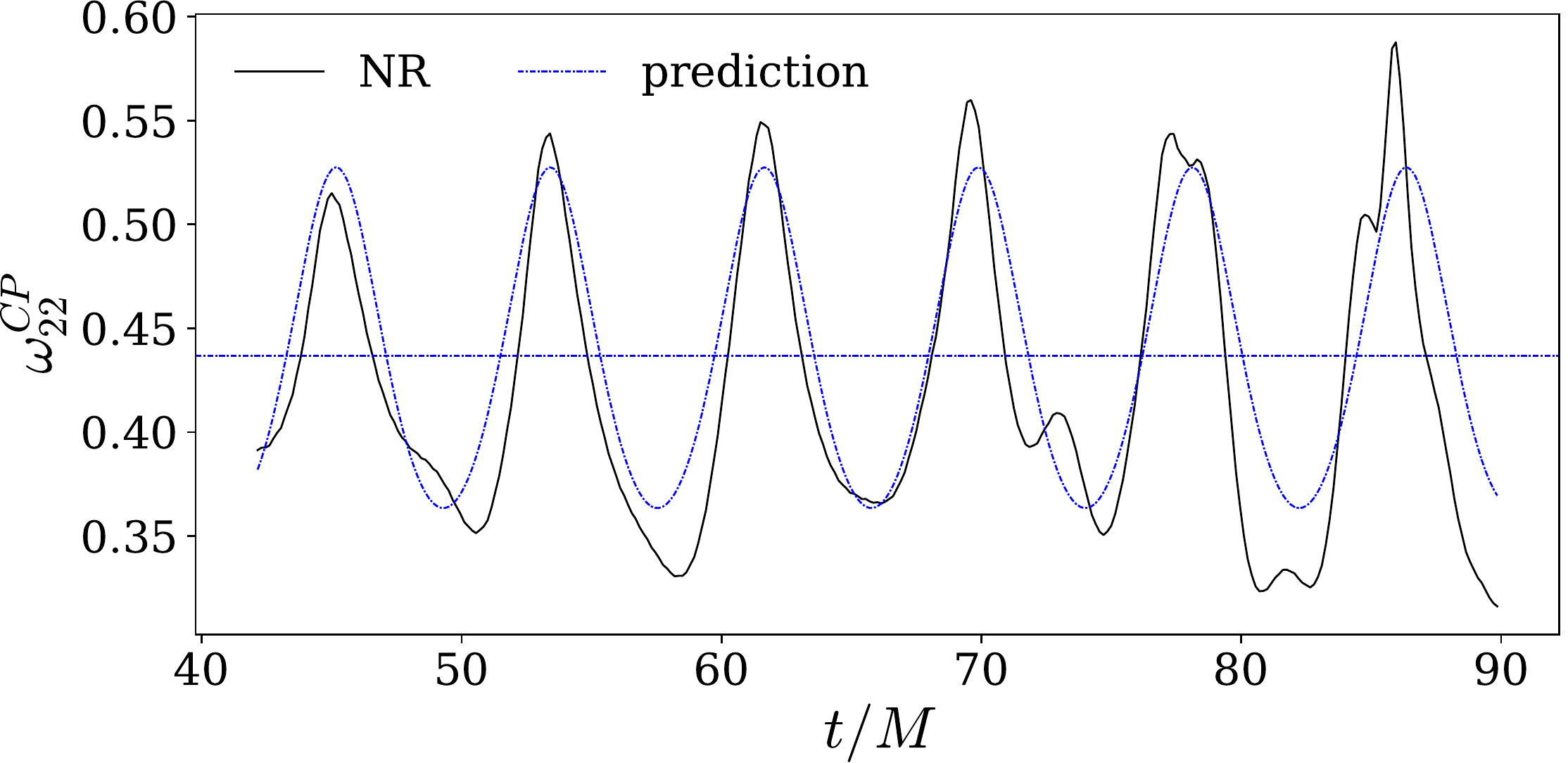} \\
   \includegraphics[width=0.48\textwidth]{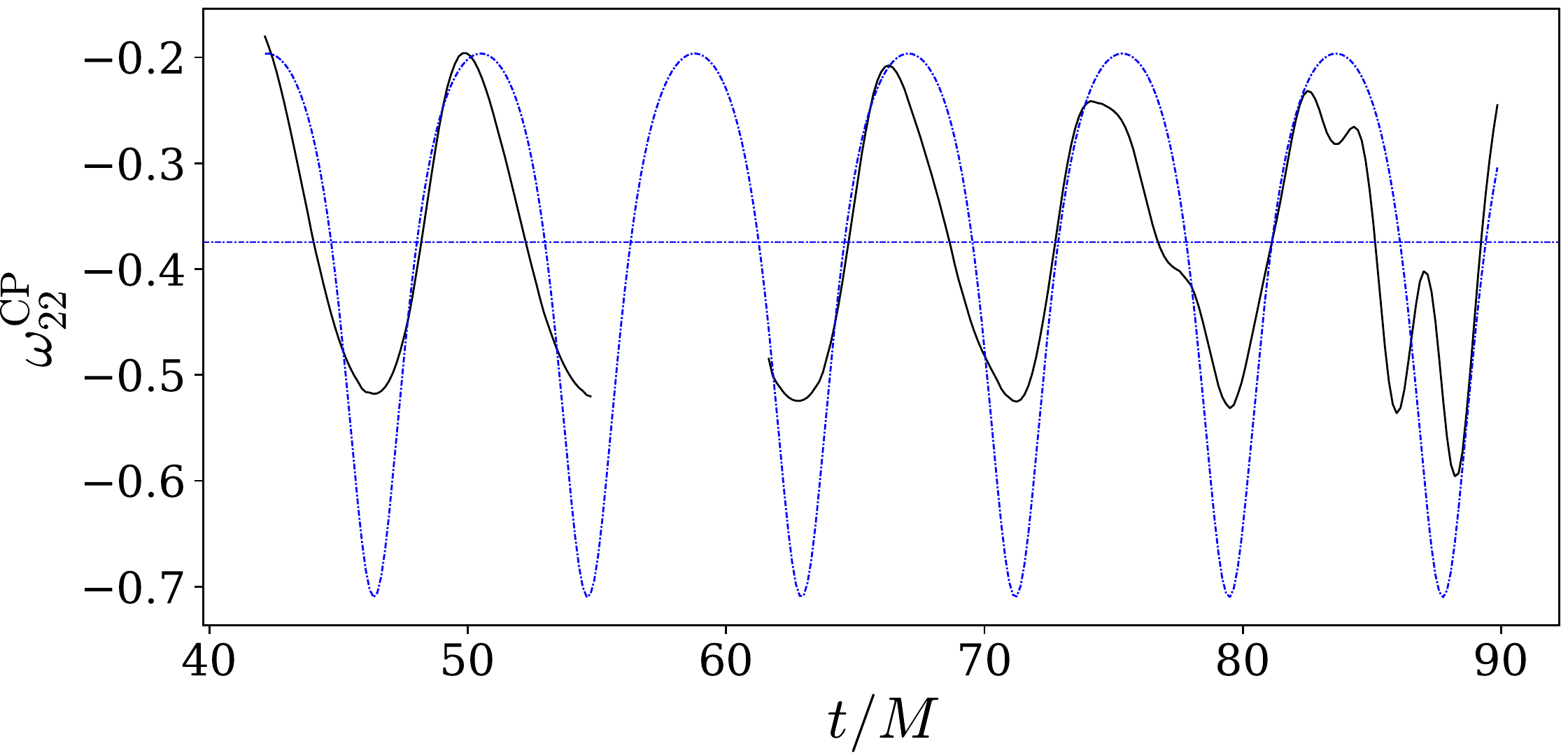}
   \caption{A comparison of the time domain angular frequency of the (2,2) multipole in the co-precessing frame with that predicted by Eq.~(\ref{eqn: full cp freq}) from the QNM picture. The horizontal line gives the prediction of the effective ringdown frequency as given by Eq.~(\ref{eqn: l2 mean beta}). The top panel shows the case $\left(q,\chi,\theta_\text{LS}\right) = \left( 8, 0.8, 120^\circ\right)$ while the bottom panel shows the case $\left(q,\chi,\theta_\text{LS}\right) = \left( 8, 0.8, 150^\circ\right)$. In the bottom case we have excised a glitch in the NR data.}
   \label{fig: oscillatory angular frequency co-precessing}
\end{figure}

The validity of these approximations in demonstrated in Fig \ref{fig: oscillatory angular frequency co-precessing}. The top panel shows a case where the prograde frequency dominates while the bottom panel shows a case where the retrograde frequency dominates. We can see that the phasing of the oscillations is consistent with $\left(2\kappa-1\right)\omega_{22}$ in both cases. The amplitude of the oscillations is captured well for the case displayed in the top panel but not in the case in the bottom panel, implying that in this case contributions from higher order multipoles in the co-precessing frame are important. Regardless of whether the amplitude of the oscillations is well captured, we can see that the expression for the mean frequency Eq.~(\ref{eqn: l2 mean beta}) accurately captures the mean behaviour showing that other than determining which \qnm{} we wish to select, the effective ringdown frequency is independent of the multipole amplitudes.

\section{Conclusions}

We have presented a simple analytic formula for calculating the effective ringdown frequency of a precessing binary in the co-precessing frame.
Such a formula is useful in accurately modelling the ringdown of precessing systems, and will be used in the model presented in Refs.~\cite{London:2023,Thompson:2023}.
It relies only on the \qnm{} frequencies (known from perturbation theory) and the value of the precession angle $\beta$ in the ringdown. 

We have also gained an increased understanding of the ringdown signal from precessing binaries.
We have seen that both the prograde and retrograde frequencies can be excited  to an approximately equal degree in these systems, resulting in oscillations in the angular frequency of the individual multipoles.
These oscillations are present in our numerical data at exactly the frequency predicted from perturbation theory, giving support to the fact that they arise from the excitation of both the prograde and retrograde frequencies.
We can obtain an independent estimate of the ratio of the amplitudes of these two different contributions from fits to the angular frequency.
Conversely, given the value of the ratio of the amplitudes of these two contributions we can provide a simple model of the oscillations in the frequency.
Such a simple means of modelling the oscillations may be useful in time-domain modelling of the ringdown.

Additionally, we have seen that the dominant ringdown behaviour is described by a single frequency, given by the average of the oscillations.
This mean value is equal to either the prograde or the retrograde frequency, depending on which dominates.
There is therefore a discontinuous transition from the prograde to the retrograde value at the point where the contributions from the two components are equal.
We find that this transition point is most consistent with where the angle between the optimal emission direction and the final spin (given by $\beta$) is equal to $\frac{\pi}{2}$.

By contrast, in the co-precessing frame we no longer have a sharp transition between the prograde and retrograde frequencies.
The smooth behaviour of the mean frequency is well captured by the analytic formula we present for the effective ringdown frequency in 
Eq.~(\ref{eqn: HOM expression}). The effectiveness of this formula in predicting the effective ringdown frequency in the co-precessing frame
for a range of multipoles is shown in Figs.~\ref{fig: RD beta comparison} and \ref{fig: 32 RD beta comparison}. 
This smooth behaviour of the analytic formula also supports our conclusion that the transition from the signal being prograde-dominated to retrograde-dominated occurs at $\beta = \frac{\pi}{2}$.

In this paper, we have therefore provided a comprehensive description of the behaviour of the ringdown frequency of precessing systems in both the inertial $\mathbf{J}$-frame and the co-precessing frame. 
This description will be applicable to future efforts in both time- and frequency-domain modelling of precessing binaries.
Accurately capturing this ringdown behaviour will also have important implications for parameter estimation of detected signals and tests of general relativity using gravitational waves.

\section{Acknowledgements}

We would like to thank Edward Fauchon-Jones, Charlie Hoy, Chinmay Kalaghatgi, Dave Yeeles, Shrobana Ghosh, Sebastian Khan, Panagiota Kolitsidou and Alex Vano-Vinuales for their efforts in running the catalogue of simulations on which this paper is based.
We also thank Jonathan Thompson for many useful discussions and Gregorio Carullo for useful feedback on the manuscript. 

E. Hamilton was supported in part by Swiss National Science Foundation (SNSF) grant IZCOZ0-189876. 
L. London was supported at Massachusetts Institute of Technology~(MIT) by National Science Foundation Grant No. PHY-1707549 
as well as support from MIT’s School of Science and Department of Physics; at King's College London by Royal Society University Research Grant {URF{\textbackslash}R1{\textbackslash}211451}; and at the University of Amsterdam by the GRAPPA Prize.
M. Hannam was supported in part by Science 
and Technology Facilities Council (STFC) grant ST/V00154X/1 and European Research Council (ERC) Consolidator Grant 647839.

Simulations used in this work used the DiRAC@Durham facility managed by the Institute for Computational Cosmology on behalf of the STFC DiRAC HPC Facility (www.dirac.ac.uk). The equipment was funded by BEIS capital funding via STFC capital grants ST/P002293/1, ST/R002371/1 and ST/S002502/1, Durham University and STFC operations grant ST/R000832/1. 
Additionally, this research was undertaken using the supercomputing facilities at Cardiff University operated by Advanced Research Computing at Cardiff (ARCCA) on behalf of the Cardiff Supercomputing Facility and the HPC Wales and Supercomputing Wales (SCW) projects. We acknowledge the support of the latter, which is part-funded by the European Regional Development Fund (ERDF) via the Welsh Government.

\bibliography{references.bib}

\end{document}

%% file: macro.tex
\newcommand\red[1]{{\color[rgb]{0.75,0.0,0.0} #1}}
\newcommand\green[1]{{\color[rgb]{0.0,0.60,0.08} #1}}
\newcommand\blue[1]{{\color[rgb]{0,0.20,0.65} #1}}
\newcommand\cyan[1]{{\color[HTML]{00c3ff} #1}}
\newcommand\bluey[1]{{\color[rgb]{0.11,0.20,0.4} #1}}
\newcommand\gray[1]{{\color[rgb]{0.7,0.70,0.7} #1}}
\newcommand\grey[1]{{\color[rgb]{0.7,0.70,0.7} #1}}
\newcommand\white[1]{{\color[rgb]{1,1,1} #1}}
\newcommand\darkgray[1]{{\color[rgb]{0.3,0.30,0.3} #1}}
\newcommand\orange[1]{{\color[rgb]{.86,0.24,0.08} #1}}
\newcommand\purple[1]{{\color[rgb]{0.45,0.10,0.45} #1}}
\newcommand\note[1]{\colorbox[rgb]{0.85,0.94,1}{\textcolor{black}{\textsc{\textsf{#1}}}}}
\definecolor{brown-ish}{RGB}{201,130,20}

\def\gw#1{gravitational wave#1}
\def\rd#1{ringdown#1}
\def\gr#1{general relativity (GR)#1\gdef\gr{GR}}
\def\nr#1{numerical relativity (NR)#1\gdef\nr{NR}}
\def\bh#1{black-hole#1 (BH#1)\gdef\bh{BH}}
\def\bbh#1{binary black hole#1 (BBH#1)\gdef\bbh{BBH}}
\def\Bbh#1{Binary black hole#1 (BBH#1)\gdef\bbh{BBH}}
\def\cbc#1{compact binary coalensence#1 (CBC#1)\gdef\cbc{CBC}}
\def\qa#1{quadrupole-aligned#1 (QA#1)\gdef\qa{QA}}
\def\pn#1{post-Newtonian (PN)#1\gdef\pn{PN}}
\def\qnm#1{Quasinormal Mode#1 (QNM#1)\gdef\qnm{QNM}}
\def\eob#1{effective-one-body (EOB)#1\gdef\eob{EOB}}
\def\imr#1{inspiral-merger-ringdown (IMR)#1\gdef\imr{IMR}}
\def\msa#1{multi-scale analysis (MSA)#1\gdef\msa{MSA}}

\def\Fig#1{Figure~\ref{#1}}
\def\fig#1{Fig.~\ref{#1}}
\def\cfig#1{Fig.~\ref{#1}}
\newcommand{\figs}[2]{Figures~(\ref{#1}-\ref{#2})}
\newcommand{\Figs}[2]{Figures~(\ref{#1}-\ref{#2})}
\newcommand{\Figsa}[2]{Figures~(\ref{#1}) and (\ref{#2})}
\def\Eqn#1{Equation~(\ref{#1})}
\def\eqn#1{Eq.~(\ref{#1})}
\def\ceqn#1{Eq.~\ref{#1}}
\newcommand{\Eqns}[2]{Equations~(\ref{#1}-\ref{#2})}
\newcommand{\Eqnsa}[2]{Equations~(\ref{#1}) and (\ref{#2})}
\newcommand{\eqns}[2]{Eqs.~(\ref{#1}-\ref{#2})}
\newcommand{\eqnsa}[2]{Eqs.~(\ref{#1}) and (\ref{#2})}
\newcommand{\ceqns}[2]{Eqs.~\ref{#1}-\ref{#2}}
\newcommand{\ceqnsa}[2]{Eqs.~\ref{#1} and \ref{#2}}